\documentclass[acmlarge]{acmart}

\setcopyright{acmlicensed}
\copyrightyear{2025}
\acmYear{2025}
\acmDOI{10.1145/XXXXXXX.XXXXXXX}

\acmJournal{IMWUT}
\acmVolume{9}
\acmNumber{3}      
\acmYear{2025}
\acmMonth{9}       

\AtBeginDocument{%
  \providecommand\BibTeX{{\normalfont B\kern-0.1em{\scshape i\kern-0.15em b}%
  \kern-0.08em\TeX}}}

\usepackage{graphicx}
\usepackage{subcaption}
\usepackage{multirow}
\usepackage{algorithm,algorithmic}
\usepackage{xcolor}
\usepackage{csquotes}   
\usepackage{enumitem}
\usepackage{textcomp}
\usepackage{svg}
\usepackage{array}

\graphicspath{{figures/}}

\begin{document}

\title[AudioMiXR]{AudioMiXR: Spatial Audio Object Manipulation
with 6DoF for Sound Design in Augmented Reality}

\author{Brandon Woodard}
\authornote{Both authors contributed equally to this research. Work primarily done at Dolby Laboratories.}
\affiliation{\institution{Brown University}\country{United States}}
\email{brandon\_woodard@brown.edu}
\orcid{0000-0003-4667-656X}

\author{Margarita Geleta}
\authornotemark[1]
\affiliation{\institution{University of California at Berkeley}\country{United States}}
\email{geleta@berkeley.edu}
\orcid{0000-0001-5823-9776}

\author{Joseph J.\ LaViola Jr.}
\affiliation{\institution{University of Central Florida}\country{United States}}
\email{jjl@cs.ucf.edu}
\orcid{0000-0003-1186-4130}

\author{Andrea Fanelli}
\affiliation{\institution{Dolby Laboratories}\country{United States}}
\email{andrea.fanelli@dolby.com}
\orcid{0009-0008-4349-2371}

\author{Rhonda Wilson}
\affiliation{\institution{Dolby Laboratories}\country{United States}}
\email{rhonda.wilson@dolby.com}
\orcid{0009-0007-3262-4845}

\renewcommand{\shortauthors}{Woodard et al.}

\begin{abstract}
 We present AudioMiXR, an augmented reality (AR) interface intended to assess how users manipulate virtual audio objects situated in their physical space using six degrees of freedom (6DoF) deployed on a head-mounted display (Apple Vision Pro) for 3D sound design. Existing tools for 3D sound design are typically constrained to desktop displays, which may limit spatial awareness of mixing within the execution environment. Utilizing an XR HMD to create soundscapes may provide a real-time test environment for 3D sound design, as modern HMDs can provide precise spatial localization assisted by cross-modal interactions. However,  there is no research on design guidelines specific to sound design with 6DoF in XR. To provide a first step toward identifying design-related research directions in this space, we conducted an exploratory study where we recruited 27 participants, consisting of expert and non-expert sound designers. The goal was to assess design lessons that can be used to inform future research venues in 3D sound design. We ran a within-subjects study where users designed both a music and cinematic soundscapes. After thematically analyzing participant data, we constructed two design lessons: (1) Proprioception for AR Sound Design, and (2) Balancing Audio-Visual Modalities in AR GUIs. Additionally, we provide application domains that can benefit most from 6DoF sound design based on our results. To expand on these insights, we conducted a second within-subjects study comparing AudioMiXR to a 2D panner baseline. Results show that AudioMiXR significantly improved usability (SUS), reduced frustration and mental workload (NASA-TLX), and enhanced creativity across all subscales. These findings demonstrate that 6DoF AR interaction not only introduces novel experiential affordances, but also yields measurable gains in user experience and creative output, positioning AudioMiXR as a promising foundation for future AR-based sound design tools.\end{abstract}

\ccsdesc[500]{Human-centered computing~Mixed / augmented reality}
\ccsdesc[500]{Human-centered computing~Auditory feedback}
\ccsdesc[500]{Human-centered computing~Gestural input}

\keywords{Augmented Reality, Spatial Audio, Sound Design, Interactive Technologies}
\begin{teaserfigure}
  \centering
  \includegraphics[width=0.75\textwidth]{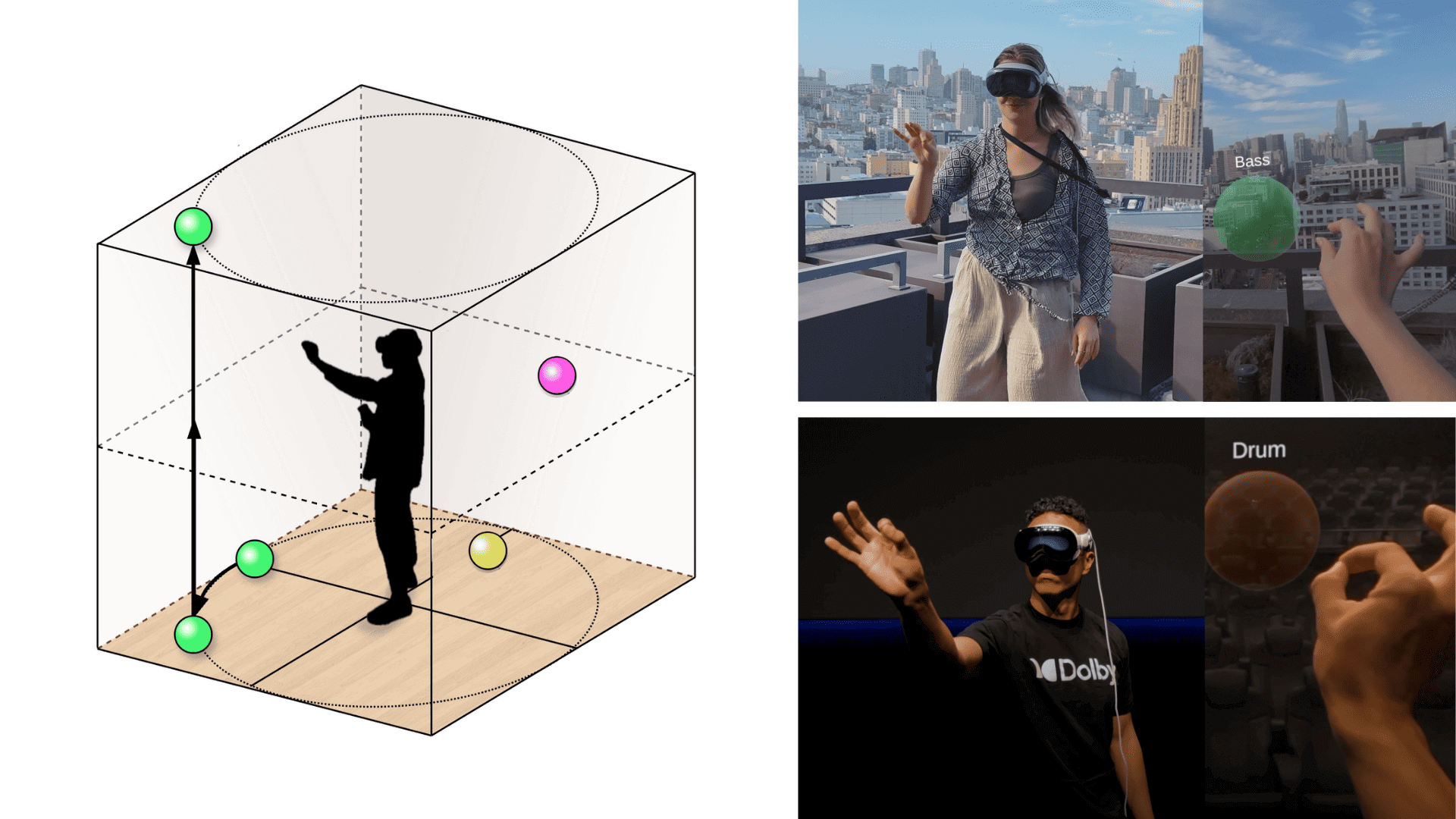}
  \caption{(Left) A schematic illustrating a user surrounded by virtual audio
  objects in 3D space. The arrows represent the virtual‑audio‑object panning
  initiated by the user’s gesture; (Right) Real‑world images of AudioMiXR users
  employing directed free‑hand manipulation for music and cinematic sound
  design, respectively.}
  \label{fig:teaser}
\end{teaserfigure}
\maketitle            

\section{INTRODUCTION}
 
Traditional approaches to spatial audio, such as channel-based surround sound systems, have been gradually supplanted by the more versatile and immersive object-based audio format \cite{sinclair2020principles, mathew:hal-01517188}. This format facilitates placement of each sound source as independent audio objects within a 3D space. Although existing tools for spatial audio panning within audio mixing frameworks provide more precise and flexible spatialization, they remain confined to 2D user interfaces \cite{marrington2017composing, misha2023}. This limitation constrains the creative freedom needed for immersive mixes and spatialized audio editing, and the transition towards the object-based audio paradigm has introduced implementation complexities that remain unresolved.
One such challenge lies in the sequential nature of 3D panning. Often, movement along the $z$-axis is constrained after the initial adjustments in the $x$-$y$ plane, limiting the creative flexibility and the natural coherence of spatial audio compositions. 
Moreover, the complexity of user interfaces for spatial audio panners can result in a steep learning curve. Further, binaural rendering can complicate audio object localization within mixing environments due to the translation between 3D user interfaces (3DUI) rendered on desktop displays and real-world applications \cite{jens1996binaural}.
 
Currently, most commercial Digital Audio Workstations (DAWs) and sound design software still rely on planar controls to manipulate the placement and motion of 3D audio objects \cite{marrington2017composing}. 
This setup forces audio engineers and producers to mentally map 3D sound onto 2D screens, often resulting in higher cognitive load \cite{farnell2010designing}. Extended reality (XR) -- comprising both augmented reality (AR) and virtual reality (VR) -- offers an opportunity to overcome these limitations, especially when six degrees of freedom (6DoF) is employed. Within a 6DoF XR interface, users can move freely in real or virtual space (translation and rotation) rather than just relying on purely rotational or \enquote{point-and-click} interactions. Such physical freedom has been shown to facilitate more accurate cognitive map building \cite{walking2011, Pastor2024}, and can enable direct manipulation  of virtual audio objects situated within the user's environment. 


Motivated by the challenges identified through conversations with professional audio engineers -- including the difficulty of creating immersive mixes, localizing binaurally rendered sound sources, and mastering steep learning curves for 3D panners -- we developed AudioMiXR, an XR-based user interface for sound design and audio mixing, designed for XR-capable head-mounted displays (HMDs), offering free-hand gestural interactions with visually rendered 3D audio objects. While it is focused on AR HMDs in its current implementation, it can be adapted to VR contexts. By leveraging how the human perceptual system understands spatial relationships, AudioMiXR aims to make immersive sound design and audio mixing more intuitive, and reduce the cognitive load of 3D panning.

This premise leads to our research questions, which seek to quantify and characterize the benefits of spatial audio interaction techniques 
in AR:
\begin{enumerate}[label=RQ\arabic*.]
    \item How should we design an XR sound design interface that leverages 6DoF? 
    \item What application areas will benefit most from an AR sound design interface with 6DoF? 
\end{enumerate}

\section{RELATED WORK}

We present an overview of relevant concepts and prior work in 3D sound design and rendering, spatial localization in XR and the utilization of expanded degrees of freedom, a brief outline of sonified experiences in XR, and XR tools for sound design and audio mixing.

\subsection{3D Sound Design and Rendering}

Modern audio production is frequently carried out within DAWs -- software environments for music creation \cite{marrington2017composing}. Popular DAWs such as Cubase, Nuendo, Logic Pro, Ableton, and Pro Tools provide visual interfaces for manipulating 
digital audio, often adopting a skeumorphic design to replicate the look and feel of legacy hardware (e.g., mixers, drum machines, or synthesizers). While this approach helps users transition from traditional studio setups by mimicking familiar controls and interaction paradigms, it can constrain the interfaces to 2D representations. Furthermore, most DAWs rely on a timeline-based structure which encourages the sequential arrangement of audio objects \cite{marrington2017composing}. This temporal editing view does not take into account the \enquote{spatial} aspect of audio. Under the \emph{object-based audio} paradigm, each sound source (or \enquote{object}) is treated as an independent entity accompanied by metadata describing its spatial attributes -- namely, the location of \emph{spatial audio} in a 3D space \cite{presence1997, begault19943}. For spatial editing, DAWs offer 3D panners, which are tools to control the position of each audio object in a 3D space \cite{mathew:hal-01517188, apple2024logicpro}. A prime example is the Dolby Atmos Panner available in Pro Tools. 

There are several approaches to spatial audio rendering, and we can highlight two in the context of object-based audio workflows \cite{sinclair2020principles}. The first is \emph{loudspeaker stereophony}, which consists in placing multiple speakers around the listener to create a surround sound experience commonly used in cinemas, home theaters, and gaming. A limitation of loudspeaker-based setups is that the listener must typically remain in an optimal \enquote{sweet spot} to fully appreciate the intended spatial effect, and phantom sound sources (i.e., perceptual illusions of sound sources) are restricted to positions between the speakers. The second approach, thanks to which many of audio-augmented XR applications are possible, is \emph{binaural rendering}. This method reproduces 3D audio using only two channels, simulating how sound is perceived by the human auditory system \cite{jens1996binaural, mauro2013binaural}. By convolving audio signals with head-related transfer functions (HRTFs), which model how sound interacts with anthropometric features, binaural rendering can simulate the localization, timbre, and externalization of sound sources, creating a realistic perception of spatial objects in a 3D environment.
However, this technique can suffer from perceptual errors, such as front-back or up-down confusions and \enquote{in-head localization} effects \cite{yang2022audio}. 

Once a spatial mix is finalized, the sequence of objects is sent to the renderer -- for example, the Dolby Atmos Renderer (potentially configured with personalized HRTFs). The renderer interprets the scene metadata to produce the final output for various configurations, whether loudspeaker arrays (e.g., 5.1.4, 7.1.4, 22.1) or headphone-based binaural playback. 

\subsection{Spatial Localization with Expanded Degrees of Freedom in XR}

XR spans both augmented reality (AR) and virtual reality (VR), combining the physical world with a digital twin world in an interactable environment \cite{milgram1995augmented}. A central factor in XR experiences is the notion of \emph{degrees of freedom} (DoF), which refers to the number of independent ways users can move within these virtual or augmented environments, and consists of translational and rotational components. Per convention, 3DoF restricts head tracking to the three rotational axes (pitch, yaw, roll), allowing users to look around but not physically move in space. A more immersive XR experience comes from adding to the equation the translational component (right/left, up/down, forward/back), resulting in 6DoF movement, which enables a more immersive and naturalistic exploration and navigation. An essential component for a user's navigation in an XR environment is building the cognitive map -- an internal representation of locations within a world-reference frame. Prior work has shown that 6DoF movement develops a more accurate cognitive map compared to 3DoF or purely joystick-based navigation \cite{walking2011, Pastor2024}. When users can physically walk around, rather than just pivot in place or point a controller, they gain a richer \enquote{movement fidelity} of real-world locomotion. 

Beyond navigation, 6DoF also permits precise placement of virtual objects. Instead of pointing a device (e.g., controller) or hand gesture in 3DoF to interact with graphical user interfaces (GUIs) situated spatially around them, with 6DoF, users can physically move to the target location and place those virtual directly \cite{kari2023scene}. This approach magnifies the realism of XR soundscapes in 6DoF audio experiences for storytelling or entertainment. 
For instance, many 360 degree cinematic experiences are in 3DoF and allow the user to look around the scene as if they were also co-located with the actors. Video games involving 6DoF interactions facilitate realism due to the envelopment of a user's entire body as an input device to interact with the content, encouraging direct engagement with XR content \cite{belonging2024}.  

Spatial audio is central to audio augmentation, such that virtual sounds are perceived as emanating from specific locations in 3D space \cite{Ruminski2015}. Humans rely on multiple sensory modalities when they engage with their environment \cite{new_ears_2024, metatla2016tap, kari2023scene}, and the auditory sense remains highly significant for localization even when visual cues is limited, sometimes replacing visual information altogether (e.g., \enquote{watching} television from another room) \cite{correa2023spatial, nonspeech1994}. Research indicates that the use of spatial audio encourages users to adopt a more active role in spatial navigation, leading to even more accurate cognitive maps \cite{Clemenson2021}, while simultaneously reinforcing the sense of presence in XR environments \cite{Kern2020-nh}. As reported in \cite{new_ears_2024}, audio interaction techniques should present directional and distance cues in contextually meaningful ways. Finally, an enhanced feeling of immersion often requires not just \enquote{looking and hearing around}, but also \enquote{moving around} in 6DoF to achieve the full benefits of spatial localization \cite{farnell2010designing}.

\subsection{Sound Experiences in XR}
Sound experiences in XR focus on interactions that leverage multimodal GUIs afforded by AR or VR head-mounted displays (HMDs), providing audio experiences that are more expressive and facilitate engagement for users. A notable example is \emph{Spatial Orchestra}, an AR interface, presented by \citet{kim2024spatial} which allowed users to walk into \enquote{bubbles} in a fixed position and co-located near each other emitting musical notes, that can only be heard once they were fully inside the bubble. This provided users with an audio experience where they could interact with music with their body without having to play a traditional instrument; although in order for the interface to work, users had to remain within a confined virtual space that was 3.3 m by 3.3 m, thus limiting their translational movements. This work is one example of audiovisual experiences that have taken advantage of the spatial cues to visualize and interact with audio using methods to render 3D visualizations of sound that are reported in the literature to form stronger connections with the audio content engaging multiple senses of users due to the multimodal feedback \cite{bilbow2022evaluating,kim2024spatial,agrawal2019defining}. Other forms of sound experiences in XR include VR or AR music-based video games like \emph{Beat Saber} where users wield a saber to hit oncoming blocks containing parts of the song's beat \cite{beatSaber2024}. In this game users are able to stand up and use their body's orientation in a fixed location to interact with the game. The visual-aural feedback provided in music-based video games provide an engaging experience where musical notes tend to take a physical form engaging users in a multisensory experience. These works consist of different combinations of haptic feedback, visual representations, 3DoF, or physical interactions.  However, these approaches are experiential and do not allow for precise control of audio to compose soundscapes intended for 3D audio, for example, music, cinematic scenes, or video games. To illustrate, \cite{murphy2010spatial} and \cite{sinclair2020principles} discussed the use of spatial sound in VR for creating immersive virtual environments with a focus on computer games.

Research in audio augmented reality (AAR) has leveraged the physical environment to augment auditory feedback rendered to the user targeted \emph{navigation and location-awareness}, as spatialized \enquote{audio beacons}, verbal instructions, or guides \cite{zoo_guide_2007, audio_stickies_2013}, studies have underscored its potential for diverse domains, including  \emph{presentation and display}, particularly in museums or archaeological sites that use audio augmentations to deliver contextual information about cultural artifacts \cite{Hatala2005, archeological2012}, or increase engagement with visitors at art exhibitions through spatialized sonic artwork like \emph{Sonic Sculpture} \cite{martin2020sonic}. 

\subsection{XR tools for Sound Design and Audio Mixing}
\begin{figure}[h]

  \centering
  \includegraphics[width=\textwidth]{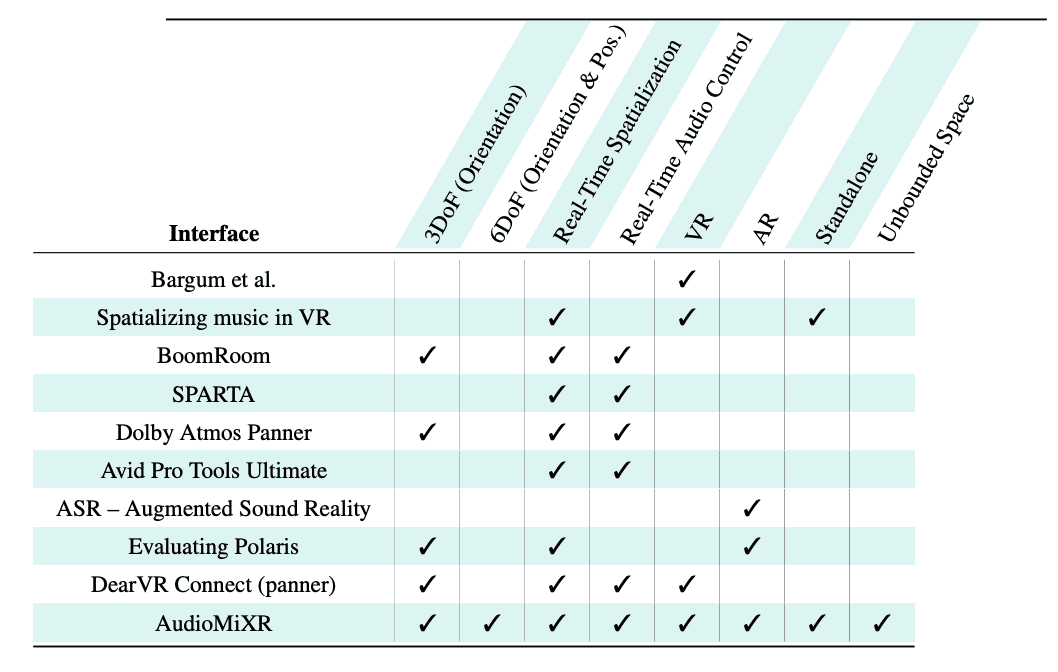}  

  \vspace{-0.5cm}
  
  \caption{Structured comparison of AudioMiXR versus traditional 2D audio tools across key dimensions. AudioMiXR combines 6DoF spatial manipulation, 6DoF binaural rendering with respect to the user's position and orientation real‐time audio control, and immersive unbounded environments. These features are absent or more limited in desktop‐based workflows or previous AR/VR interfaces intended for mixing spatial audio. Interfaces (from top to bottom): \cite{bargum2023spatial,jiang2023spatializing, muller2014boomroom, sparta2025, dolby2018atmosrenderer,avid_pro_tools, haller2002sketch, bilbow2022evaluating,dearvr2024}}
  \label{fig:comparison}
  
\end{figure}

 A subset of design tools in XR utilize the flexibility of a spatial interface to interact with floating GUIs comprising of drag-and-drop UIs that can be manipulated with controllers. A notable commercial example is \emph{DearVR}, which uses virtual knobs and faders with a similar appearance of a music production studio \cite{dearvr2024}. Several research works in sound design \cite{jiang2023spatializing,bargum2023spatial} have utilized a VR headset to recreate a virtual mixing environment, similar to the panning tools in DAWs, where users can control audio objects positioned at a distance with a controller and raycast to select the objects while remaining stationary. 
 While all of these tools utilize some spatial aspects applied to sound design, they rely on a stationary user position and none of them exploit 6DoF direct manipulation, in which users can physically move around the environment to locate and place virtual audio objects. 
By contrast, AudioMiXR integrates 6DoF direct free-hand manipulation into the audio mixing workflow, allowing the user to traverse a real or virtual environment, directly grab virtual audio objects, and reposition them wherever they see fit -- this way, enabling true \emph{hands-on} editing of spatial sound as if the users were \enquote{inside the panner}, in ways conventional DAW panners or existing VR mixing tools cannot as seen in Figure \ref{fig:comparison}.

\section{DESIGN CONSIDERATIONS}

To inform the early design of AudioMiXR, we conducted semi-structured formative interviews with four professional audio mixers 
(E1, E2, E3, E4), and followed sound design principles \cite{farnell2010designing} and AR design guidelines \cite{ar_design_heuristics_2017} to ensure AudioMiXR aligns with both established spatial audio mixing conventions and ergonomic AR design. We identify experts as people who use DAW software at a professional capacity and recruited from an industry-leading company specializing in audio technology. The insights from the interviews are listed next.

\subsection{Expert formative interview}
For our expert interviews, we adapted the structure of the formative interview method by \cite{adobe2020q}. Each one revolved around a set of 12 questions designed to uncover established spatial audio workflows, technical challenges, and preferences. These were used to design AudioMiXR by maximizing the prospective benefits of an AR interface. Refer to the appendixfor the complete list of questions.

\subsubsection{Mixing Workflow}
We began by asking experts to describe their general processes for mixing and panning spatial audio, as well as the software tools they typically use. When describing, experts emphasized how they use audio object placement in 3D to support the narrative of sonified experiences and how the visual-aural feedback is key to create a compelling experience. E2 noted that if a visual on-screen object moves, the sound should follow its motion to create a convincing illusion that they are \enquote{\emph{in sync}} -- an observation that highlights the importance of perceptual alignment between visual cues and audio sources. This notion reflects the AR design principle of \emph{alignment between physical-virtual worlds} \cite{ar_design_heuristics_2017}, ensuring that the listener's perceptual system \enquote{\emph{automatically accepts}} the audio-visual synchronization.
Experts stressed that the overall focus of a spatial mix is to create a \emph{compelling} environment -- realistic or stylized -- that meets the goal of \enquote{\emph{telling a story or supporting}} a particular user experience. In other words, the spatial mix must reflect the intended function, whether it is cinematic storytelling, gaming, or sound meditation, mirroring the AR design principle of \emph{fitting with the user environment and the task} \cite{ar_design_heuristics_2017}.

\subsubsection{Spatialization}
When asked about specific spatialization strategies, E1, E2, E3 agreed that it is ultimately \enquote{\emph{done to taste}}, although certain heuristics commonly apply. 
For instance, low-frequency elements (like bass or kick drums) often remain front and center, while higher-frequency elements might be placed in overhead channels. These practicalities resonate with how the human perceptual system associates certain frequency bands with location \cite{karla2010vision, gibson1997art}, 
such that the audio object arrangement feels coherent to the listener's mental model of the environment. In other words, it should be familiar enough so that the listener's expectations are met (i.e, \enquote{\emph{the waves should be there, the birds over there}}).

\subsubsection{Challenges of Spatial Audio}
The experts identified localization issues while mixing -- particularly front/back confusion in binaural rendering -- as a frequent challenge.
They noted that point-source audio objects can fail to convey an effective sense of immersion if the sounds remain narrowly localized or singular. 
Both E1 and E3 mentioned how changing the size or duplicating audio objects 
to occupy more of the soundscape could address this issue by providing multiple directional cues or a broader \enquote{\emph{spread}}.
Such references to \emph{spread}, \emph{duplication}, and \emph{overlapping elements} yet again reflect the sound design principle of \emph{layering} \cite{farnell2010designing}, where combining or stacking multiple sounds adds depth and complexity to the mix. It also connects with the AR principle of \emph{form communicates function} \cite{ar_design_heuristics_2017}, since manipulating the visual appearance of an audio object maps directly to certain auditory attribute changes, creating affordances.

\subsubsection{Collaborative Design}
Because spatial audio often involves interdisciplinary collaboration, we also inquired about the social and collaborative aspects of mixing. 
In current workflows, experts typically exchange binaural mixes remotely, receiving asynchronous feedback from listeners. An AR environment could potentially streamline this iterative process by allowing co-located or remote collaborators to view and hear each other's manipulations in real-time. A scenario where collaborators can be in the same room with an AR mixing interface synced with loudspeakers in the physical space or while one `operator' mixes was discussed.

\subsubsection{Future Directions}
We also asked participants about notable successes in their spatial audio practices, and future directions that might enhance their workflows. General directions pointed toward improving the UIs to be more user-friendly and intuitive and, interestingly, E3 hinted at how a more physical approach could make mixing easier and more accessible: \emph{\enquote{certainly using body movement to create and automate panning.  You could almost \enquote{dance the mix}.  This also opens up accessibility} -- E3}. This notion of using one's body movement and gestures directly touches on the applicability of AR in an audio mixing task, encouraging gestures that feel natural rather than fatiguing, referencing the AR design heuristic of \emph{fitting with user's physical abilities} \cite{ar_design_heuristics_2017}.

\subsubsection{Evaluation of a Mix}
Finally, experts reflected on how they typically evaluate the quality of a spatial audio mix, describing metrics for success. For instance, they consider whether it \enquote{\emph{sounds more interesting than [in] stereo}} or whether it has \enquote{\emph{a good sense of height, width, depth, externalization, and timbre}}. Although these mostly rely on subjective measures of audio mixer's taste, E2 noted: \emph{\enquote{the one thing that can be truly measured is the loudness}}. Generally measured in LUFS units for audio mixing applications, \emph{loudness} serves as an objective metric for ensuring that the final mix meets the delivery standards across different platforms. 

\subsection{Key Takeways}
Experts in our formative interviews revealed several core considerations for spatial audio mixing in AR which we took into account for designing AudioMiXR. Foremost, we clarified why AudioMiXR could be better than traditional 2D audio panning interfaces. Its advantage lies in the natural conversion of virtual audio objects in a 3D space, which makes it easier for creators to mold compelling experiences for diverse applications. Experts stressed that ensuring audio objects are synchronized with their corresponding visuals is crucial for immersion. Moreover, experts highlighted the benefits of a more physical, gesture-driven interface. Following on their feedback, we made sure to design AudioMiXR with real-time aural-visual feedback and to provide free-hand spatial manipulation within our interface. However, at the same time, localization challenges remain an obstacle. We leverage the audiovisual sensory component of our AR interface since humans are better at that than in unisensory localization scenarios \cite{odegaard2015biases, crossmodal2016chi}. 

Our experts also recognized the collaborative nature of spatial audio production. An AR approach could support more synchronous or alternative forms of collaboration -- allowing multiple users to see and hear each other's changes in real time. In all, the interviews point to the promise of AR-based mixing for bridging the mapping gap between the virtual spatial mix and the actual physical implementation and delivery. Taken together, these insights form the backbone of AudioMiXR. We present its detailed system design in the following section.



\section{SYSTEM}

This section describes the AudioMiXR system, an XR user interface for free-hand manipulation of spatial audio. We detail the system architecture, and provide an overview of user interaction mechanisms and the standard workflow.

\subsection{System Architecture}
To enable free-hand spatial audio manipulation, AudioMiXR relies on several hardware and software components.
On the hardware side, the system runs on an XR-capable HMD which provides a built-in device camera and depth sensors to capture the user's hand gestures and movements, producing 3D coordinates for real-time gesture recognition. Additionally, the inertial measurement unit (IMU) within the HMD tracks the position and orientation relative to a defined origin. This ensures that the system consistently aligns the user's real world head movement with the superimposed virtual environment.

On the software side, AudioMiXR is comprised of a suite of modules. First, an \emph{image tracking and gesture recognition} module processes the depth-camera feed and applies HMD's native computer vision algorithms to interpret how the user's hands move and which gestures they perform. Next, a \emph{3D mapping} module translates these hand positions into virtual coordinates, accurately tracking and registering the hand positions within a 3D coordinate space. A \emph{gesture I/O} module then assigns a class of a specific gesture to input and output functions in order to interact with the visual 3D audio objects.  To render and update these virtual audio objects, AudioMiXR has a \emph{3D simulation} module -- whenever the system detects a relevant user gesture, the 3D simulation module handles how the the visual 3D objects should respond to the users input (e.g., the user moves their hand up and down along the vertical axis then the audio object vertically translates in accordance to the hand movement and positioned at the extent of their fingertips). The \emph{communication module} updates the 3D hand coordinates and visual audio object positions for a set interval while the application is running (e.g. updates audio object position once every 30 frames via USB or wireless protocol). 
The HMD's spatial operating system, processor, and memory provide the necessary computational resources to execute computer vision algorithms, maintain stable 3D rendering, and track hand gestures while simultaneously updating audio objects.

\subsection{Deployment Target}
Although both AR and VR share common goals of augmenting user environments, they differ in key respects. 
VR often disconnects users visually from their surroundings, risking collisions or accidents while they are interacting with the digital twin world without having cues on their real environment \cite{interactions_vr_2024, vrtoer2023}. Even when using VR safety measures, such as boundaries (e.g., Meta Quest Guardian System) users may still break them, and safety comes at the expense of immersion \cite{interactions_vr_2024}. VR immersion can also induce \emph{cybersickness} -- a visual-vestibular conflict from motion in a virtual world -- which may intensify if, for instance, users rely on teleportation for locomotion \cite{new_ears_2024} or omnidirectional treadmills for infinite-walking \cite{treadmill_2024}, which even after proper training do not provide firm locomotion, but rather an awkward gait \cite{walking2011}. Ensuring perceptual-motor fidelity between a user and their digital-twin avatar can mitigate these negative effects by \enquote{virtual body ownership} \cite{hands2016, 10.1371/journal.pone.0010564}. 

Conversely, AR situates virtual content onto physical environments, allowing users to interact with virtual elements as though they were truly present in their surroundings \cite{sodnik2006, Ruminski2015, yang2022audio}, often avoiding some of VR's drawbacks, as users retain awareness of their physical environment \cite{Steffen03072019, interactions_vr_2024}. 
Given these considerations, our user study focused on the AR setting, although future extensions could address VR scenarios, since AudioMiXR runs on headset capable of supporting both AR and VR modes. While the immersive 6DoF capabilities in AR and VR are comparable in principle, VR introduces additional aforementioned caveats that might require further design adaptations.

\subsection{User Interaction and GUI}

\begin{figure}[ht]
    \centering
    \begin{subfigure}[c]{0.49\textwidth}
        \centering
        \includegraphics[width=\linewidth]{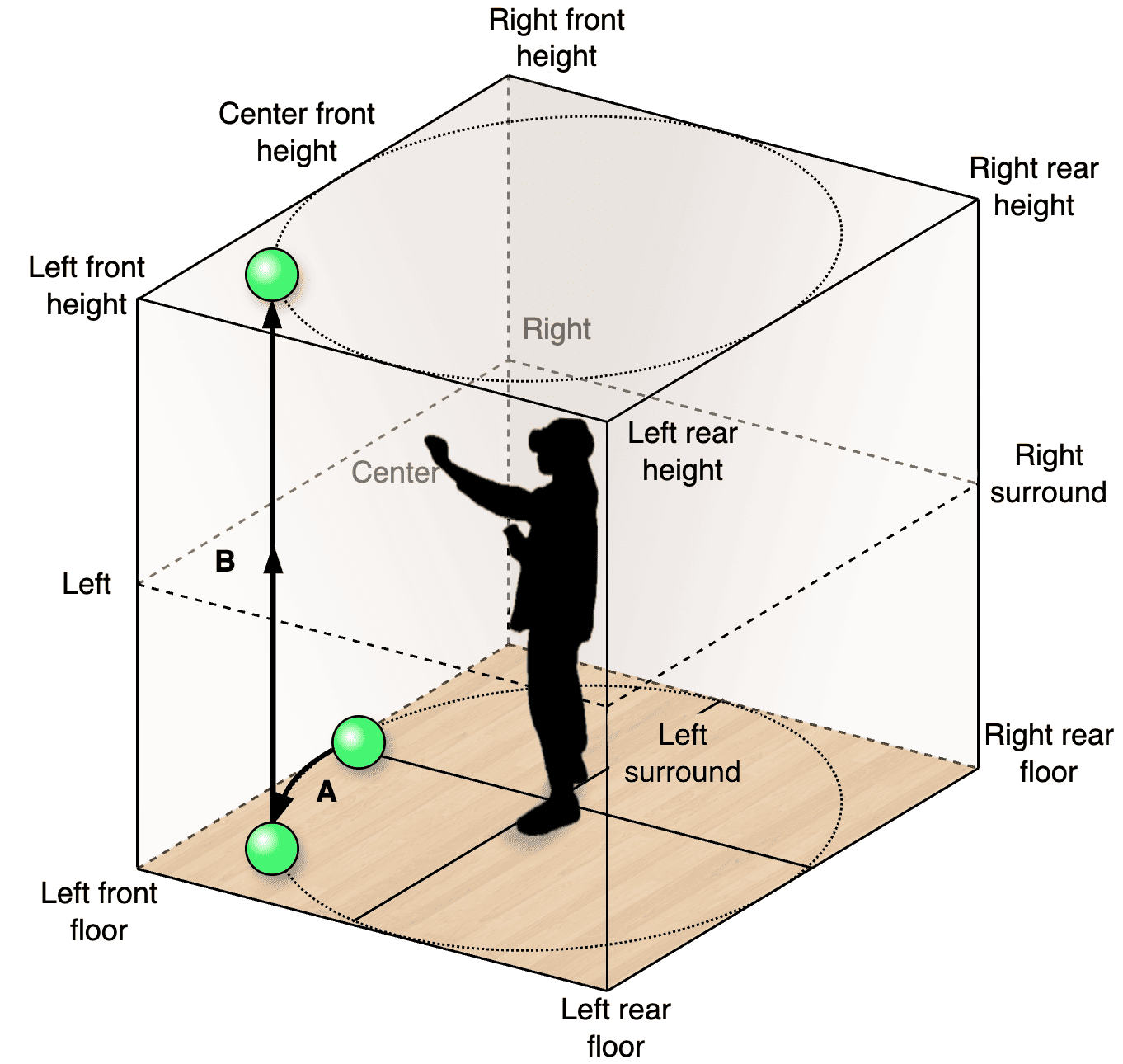}
    \end{subfigure}
    \hfill
    \begin{subfigure}[c]{0.49\textwidth}
        \centering
        \includegraphics[width=\linewidth]{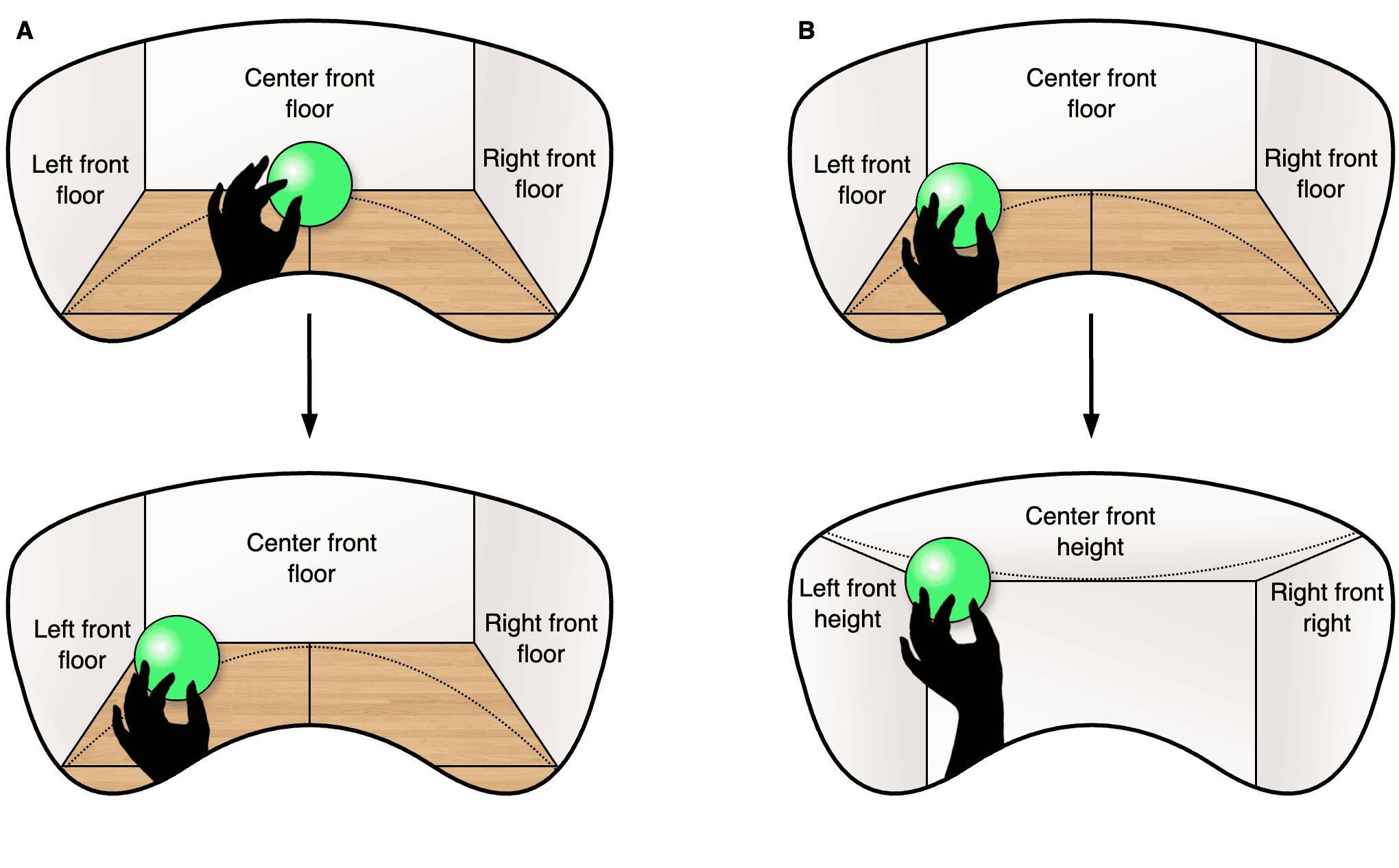}
    \end{subfigure}
    \caption{(Left-to-Right) AudioMiXR enables users to reposition audio objects within their reach envelope through direct gestural interactions, giving the impression that their hands are physically in contact with the objects. Users can push objects beyond their reach envelope by targeting the object with a pinch gesture and performing a pushing motion. If an object is out of reach and the user prefers not to physically walk towards it, they can align their pinch gesture with the object in their line of sight and \enquote{pull} it back into their reach envelope.}
    \label{fig:side-by-side}
\end{figure}

In AudioMiXR, we utilize rendered computer graphics to depict 3D audio object 
manipulations (e.g., position in space or distance from the listener). The visual representation of each sound objects is modified through a free-hand pinch gesture used for selecting and holding objects.
AudioMiXR uses simple, direct gestures to give the sensation of physically touching and/or moving virtual audio objects in the air. As shown in Figure \ref{fig:side-by-side}, users can \enquote{pinch} a virtual audio object to grab it, \enquote{push} it further away, or \enquote{pull} it closer -- even if they remain standing at a distance. 

\subsection{User Workflow}
\begin{figure}[ht!]
    \centering
    \includegraphics[width=1.0\linewidth]{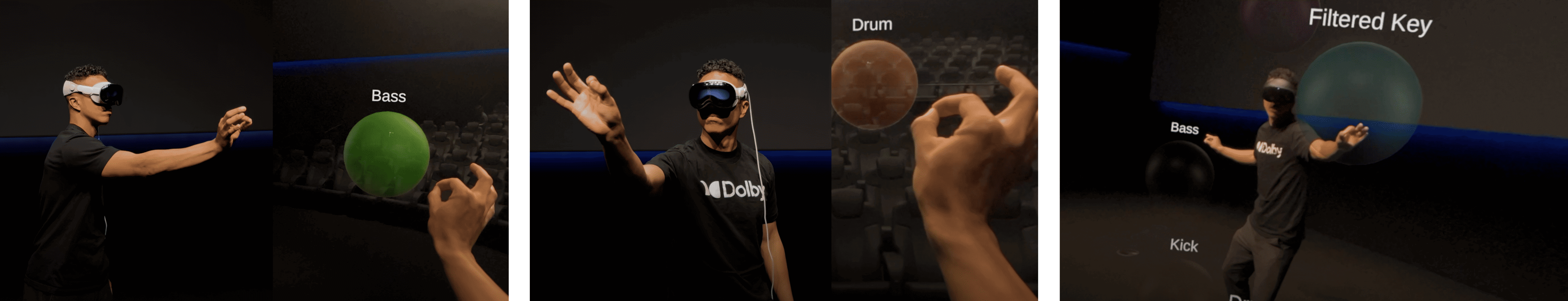}
    \caption{A user interacting with virtual audio objects in AR using direct free-hand manipulation. }
    \label{fig:user-worklow-example}
\end{figure}

Once calibration is complete, the user can start the session by booting the XR interface, which initializes the \emph{gesture recognition}, \emph{hand tracking}, and \emph{body tracking} modules. Once in the environment, the user can freely grab any virtual audio object by pinching it (Figure \ref{fig:user-worklow-example}) and reposition it within or beyond their physical reach. If the target object lies out of each, the user may \enquote{pull} it back by targeting it with the gaze, performing the pinch gesture, and making a dragging motion. Changes in position immediately translate to perceptible alterations in audio attributes.

\subsection{Implementation}
We developed AudioMiXR using the Unity game engine (runtime version 2022.3.30f1) in C\#, deploying it on an Apple Vision Pro (AVP) HMD, on its native spatial operating system visionOS 1.0. We used Unity for its robust developer community and comprehensive documentation and support for rendering and interaction. 
We used a Unity Professional license to gain access to the necessary AVP deployment modules and visionOS-specific libraries, and an Apple Developer license to sign, build, and deploy AudioMiXR to the AVP HMD.   

\begin{figure}[ht!]
    \centering
    \includegraphics[width=1.0\linewidth]{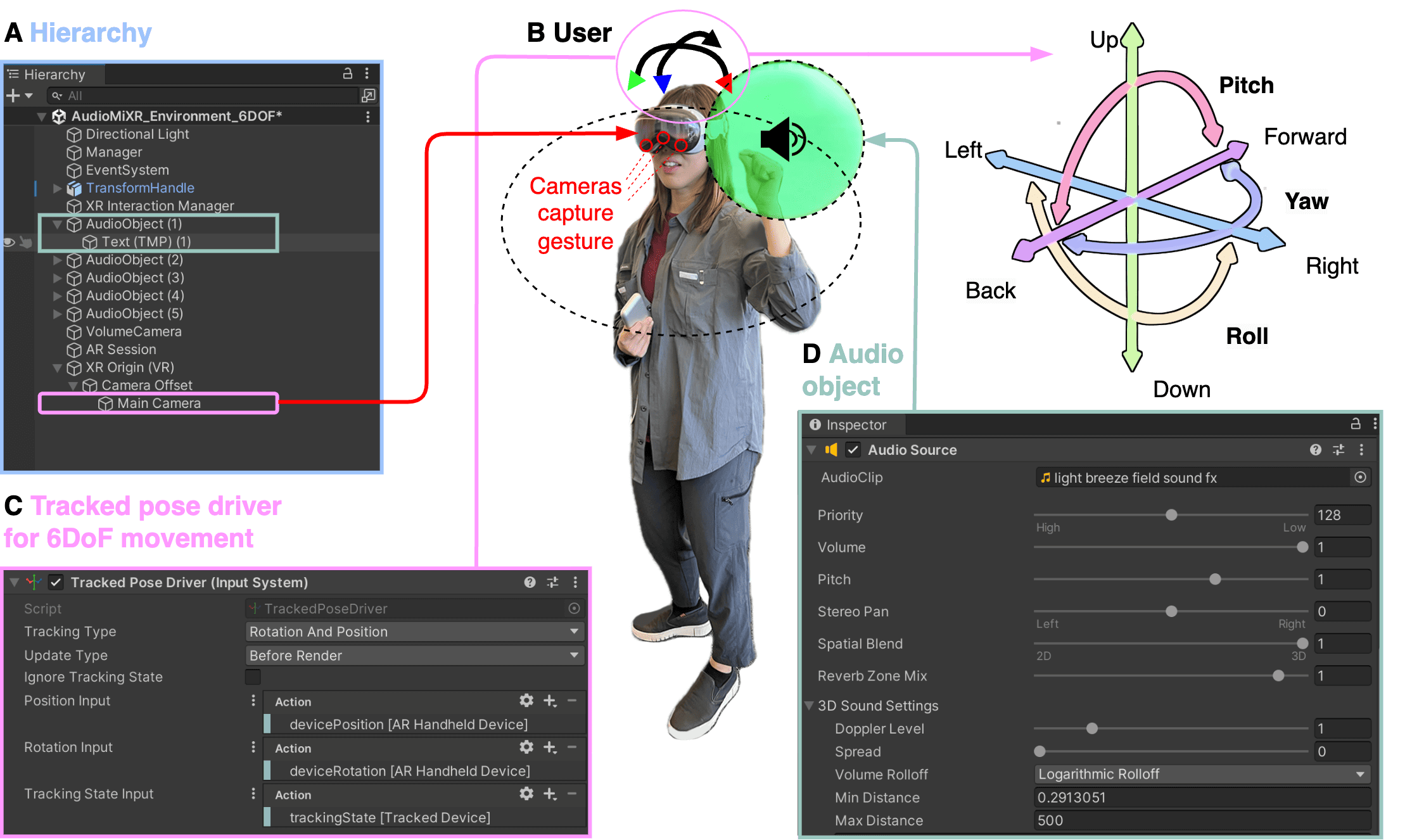}
\caption{(A) Hierarchy in the Unity3D engine showing all the game objects; (B) A user featuring the headset with a tracked pose driver enabling 6DoF movement and gesture-based interaction with a virtual audio object; (C) Configuration settings for Unity Engine's spatial tracking driver; (D) Settings of the audio source element attached to the virtual audio object. }
\label{fig:unity-ui}
\end{figure}

\subsubsection{Hardware and tracking.} The AVP provides 6DoF tracking through its built-in IMU sensors and LiDAR-based environment mapping, allowing real-time position (right/left, up/down, forward/back) and orientation (pitch, yaw, roll) updates. As illustrated in Figure \ref{fig:unity-ui}.C, we leverage Unity's \emph{Tracked Pose Driver} component on the HMD camera to enable positional and rotational tracking, configured to update sample tracking input right before the rendering step 
for minimal latency. Our \emph{XR origin} in Unity contains the audio listener -- a component that captures the incoming sound -- and references the HMD's head pose using the XR Interaction toolkit, while an additional \emph{AR session} object manages environment tracking (position and rotation) via ARKit, provided by the Unity AR Foundation library. To support free-hand spatial audio manipulation, we incorporate the AVP's integrated 3D touch input using the PolySpatial engine, allowing users to \enquote{pinch} virtual objects and reposition them in 3D space when recognized by the AVP's cameras (Figure \ref{fig:unity-ui}.B) . 
For audio playback, users can opt for AVP's native near-ear headphones or Bluetooth-enabled headphones. 




\subsubsection{Audio objects and interaction.} 
\begin{figure}[ht!]
    \centering
    \includegraphics[width=1.0\linewidth]{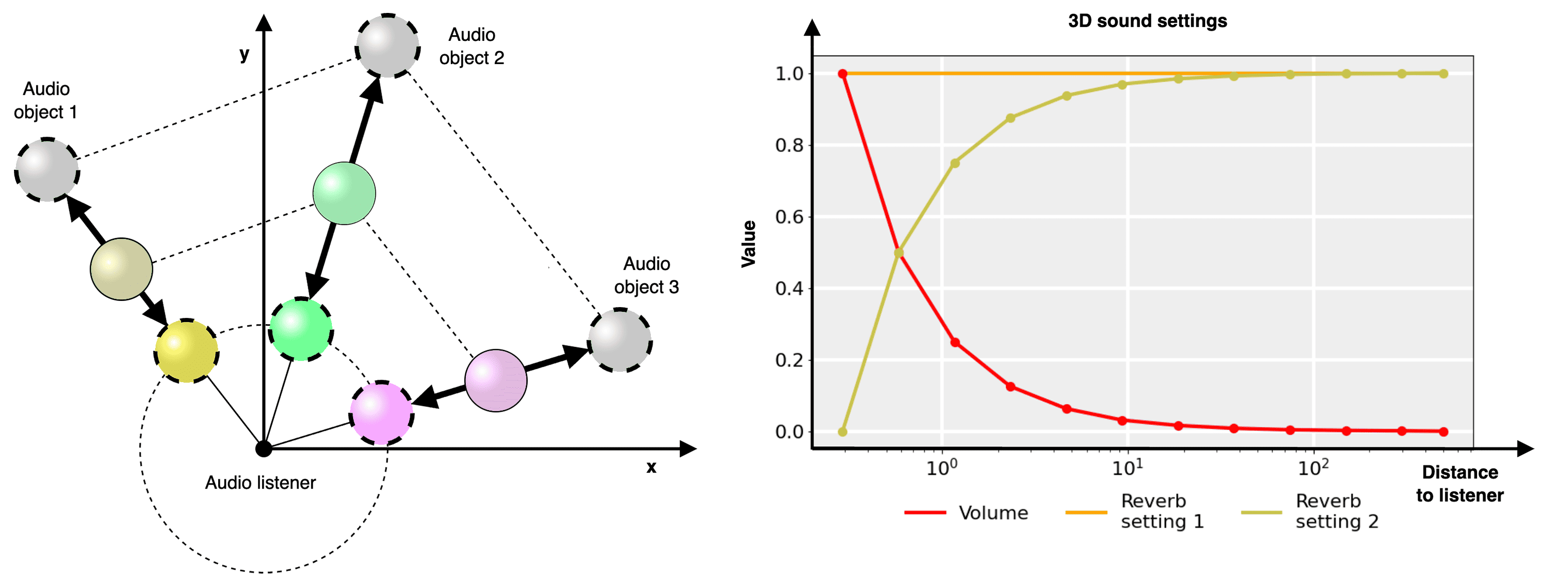}
    \caption{(Left) Schematic representation of the AR panner. The spheres represent virtual audio objects, and the color saturation represents the level decay based on distance from the audio listener located at the HMD to the audio spheres. The $X$-$Y$ axes quantify the amount of room response in the binaural rendering setup; (Right) The plot displays the 3D sound settings in the game engine. The virtual audio object level displays a logarithmic decay based on distance from the audio listener. For the reverb setting, we experimented with a uniform, full-strength reverberation (\enquote{Reverb setting 1}, our default setting for the user study) and a monotonically increasing reverb mode that adds more ambience as the distance from the listener grows (\enquote{Reverb setting 2}).}
    \label{fig:AR-panner-reverb-settings}
\end{figure}

Each audio object in our scene (Figure \ref{fig:unity-ui}.A) is represented as a spherical \emph{GameObject} Unity's fundamental object type, with the following added components: a Unity \emph{audio source} component (Figure \ref{fig:unity-ui}.D) configured for 3D spatialization with logarithmic rolloff for distance-based attenuation; a custom \emph{opacity modulation script} that adjusts visual transparency relative to the distance from the \emph{XR origin} (see left subfigure in Figure \ref{fig:AR-panner-reverb-settings}); a custom \emph{pulse-transferring script} employing a Fast Fourier Transform (FFT) with a Blackman window to sample the audio signal in real time. We segment the FFT result into frequency bands and dynamically scale each sphere based on the amplitude of the selected band (e.g., high-frequency kicks cause the sphere to pulsate, providing immediate visual feedback). 
Our scenes feature multiple audio objects representing distinct sound samples. To accomodate large spatial layouts, we set Unity's \emph{volume camera} configuration to \enquote{unbounded}, giving users freedom to walk around and place objects anywhere in their physical environment. Unity's PolySpatial engine also takes care of converting Unity's materials and mesh renderers into AVP's native rendering.

\subsubsection{Binaural rendering and reverb.} Unity's native audio engine allows binaural rendering by applying 3D sound settings (Figure \ref{fig:unity-ui}.D) that simulate how audio signals reach each ear from different directions and distances. While additional proprietary plugins exist for HRTF modeling, AudioMiXR relies on Unity's built-in audio spatializer.
For that reason, we manually deactivated AVP's native spatializer. We also experimented with two reverb schemes for the audio objects. The first, \emph{constant reverb}, 
applies a uniform reverberation effect at full strength across all distances between the virtual audio object and the audio listener, creating the sensation of a large, uniformly reverberant area. This setting served as the default option during our user studies. As an alternative, we used the \emph{distance-based reverb} mode in our supplementary demos, which starts with no reverb at close range and increases logarithmically with distance. This approach provides an effect of ambiance and space as objects move farther away from the listener. The right subfigure in Figure \ref{fig:AR-panner-reverb-settings} illustrates both the logarithmic distance-based attenuation and the corresponding reverberation curves.

\section{METHOD}

To evaluate AudioMiXR, we conducted two formal user studies  designed to uncover design lessons, usability issues, usability, and creative output. 
First we outline the methods for the exploratory study intended to identify user values and interaction patterns to inform design lessons; the second, A/B experiment (Study 2), comparing AudioMiXR to a 2D spatial audio panner, is described in the next subsection.

\subsection{Study 1 - Exploratory Design Lessons Experiment}

\subsubsection{Participants. } We recruited 27 participants via snowball sampling 
from Dolby Laboratories, which is typically sufficient to reach saturation of recurring themes in participant responses \cite{guest2006many,iftikhar2023together}. The sample included 17 male and 10 female participants, ranging in age from 22 to 59 ($\mu=31.22, \sigma=8.70$). Participants have diverse audio mixing  and AR backgrounds, which we used to categorize them into DAW \emph{experts} or \emph{non-experts}. Specifically, those who self-reported as \emph{unexperienced}, \emph{beginners}, or \emph{intermediate level} DAW users were considered \enquote{non-experts}, while those who self-identified as \emph{advanced} and \emph{expert} were considered \enquote{experts}, resulting in 18 non-experts and 9 experts.
This study aimed to quantitatively measure workload, observe user interaction patterns, and collect qualitative feedback on AudioMiXR's usability and potential professional applications. We used the AVP without optical inserts; 
consequently, we collected participants' glasses prescriptions and diopters, if available (for control). Of the 27 participants, 10 had perfect vision and 17 had different mild eye conditions (e.g., mild nearsightedness or astigmatism) 
but felt comfortable taking part in the experiment without vision correction.



\begin{figure}
    \centering
    \includegraphics[width=1.0\linewidth]{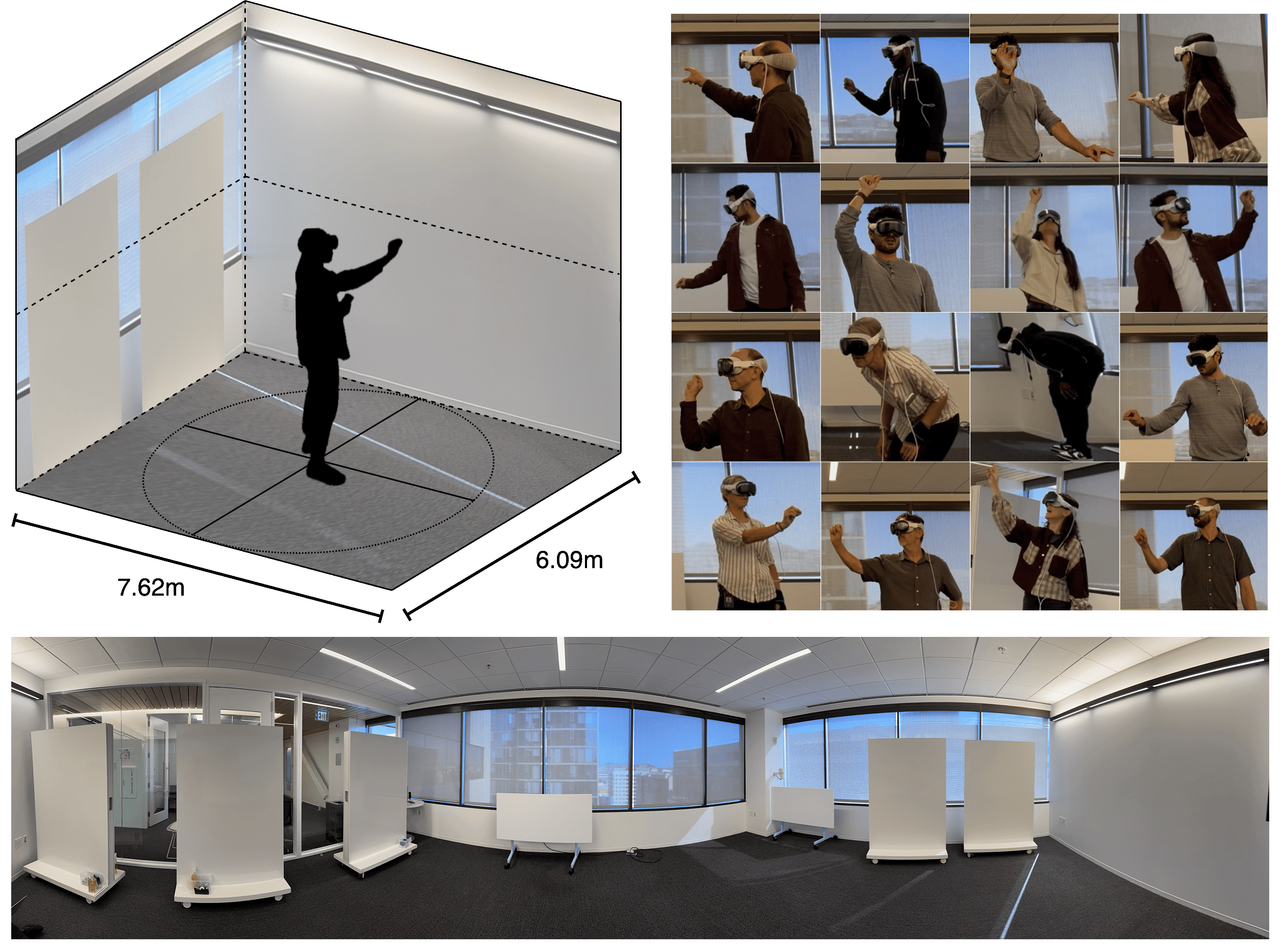}
    \caption{(Left) Schematic representation of the room where the user study was run, the dimensions of the room were $7.62$ m in width and $6.09$ m in depth. The origin was marked right at the center of the room with a white cross on the floor; (Right); Snapshots of users interacting with AudioMiXR during the user study--a variety of interesting user behaviors were capturing, including but not limited to: crouching, attempt to manipulate audio objects with two hands at once, placing audio objects over the head, etc.; (Bottom) Panoramic view of the room where the user study was run.}
    \label{fig:experiment-setup-room}
\end{figure}

\subsubsection{Setup.} The study took place in a room measuring $7.62$ m $\times 6.09$ m (width $\times$ depth), as illustrated in Figure \ref{fig:experiment-setup-room} (left). Participants wore an AVP HMD and stood at the origin point marked with a cross on the floor. We positioned an additional camera near one wall of the room to capture each participant's movement, while the AVP simultaneously recorded the first-person view of virtual audio object manipulation.

\subsubsection{Procedure.} 
\begin{figure}[ht!]
    \centering
    
    \begin{subfigure}{1.0\linewidth}
    \centering
    \includegraphics[width=1.0\linewidth]{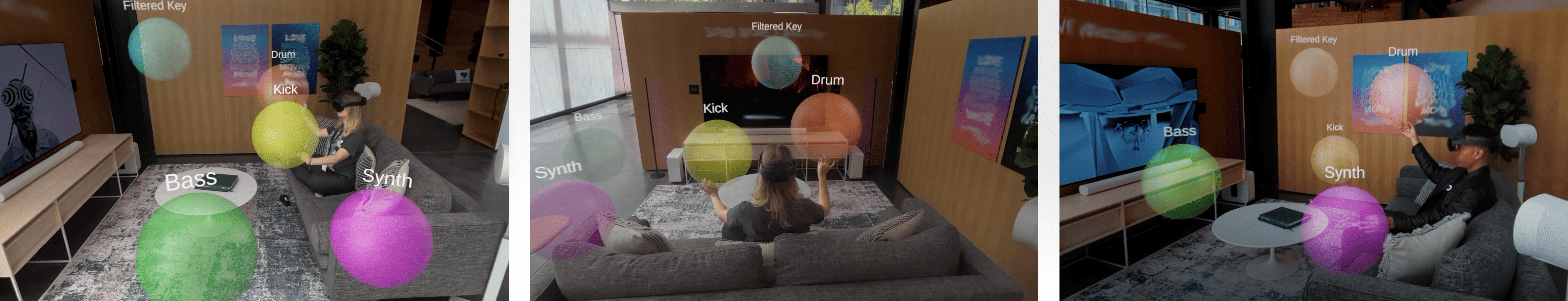}
    \end{subfigure}
    \\
    \vspace{1mm}
    \begin{subfigure}{1.0\linewidth}
    \centering
    \includegraphics[width=1.0\linewidth]{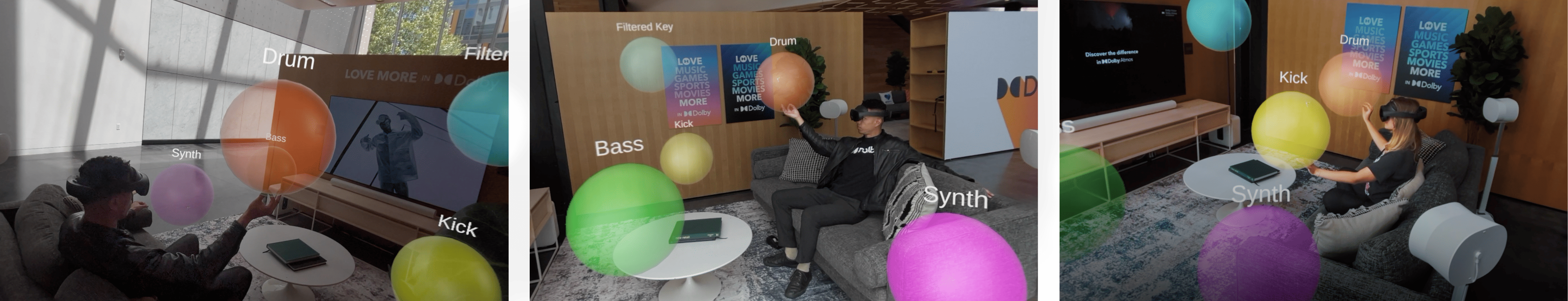}
    \end{subfigure}
    \caption{Users interacting with virtual audio objects in the music mixing scene. For illustration purposes, note that the users are using Meta Quest Pro. Since AudioMiXR has been developed in Unity3D, it can be deployed on both platforms, AVP and Meta Quest Pro.}
    \label{audio-mixing-scene-example-1}
\end{figure}

We ran a within-subjects study where each user attempted the instructed directed tasks in 6DoF. Each session began with a brief \emph{calibration phase} (~2 minutes), in which participants aligned hand-tracking and eye-tracking for the AVP, followed by a \emph{training phase} (~3 minutes) during which the participants were able to practice essential gestures -- pinching to grab and reposition virtual audio objects, walking around them (clockwise/counter-clockwise), and moving objects closer or farther away to observe changes in audio attributes. After the training session, participants were instructed to complete tasks in two distinct \emph{scenes}, each containing five unique audio objects. We choose five as a lower bound of \emph{Miller's rule} \cite{miller1956magical} -- which defines $7\pm2$ as the number of objects an average human can hold in short-term memory -- to leverage the attention span of our participants \cite{farnell2010designing}. In the \emph{music mixing scene} (Figure \ref{audio-mixing-scene-example-1}), users placed and arranged objects labeled \enquote{\emph{Synth}}, \enquote{\emph{Bass}}, \enquote{\emph{Kick}}, \enquote{\emph{Filtered Key}}, and \enquote{\emph{Drum}} to construct what they feel is the \emph{most} immersive listening experience. We described ``immersive" to participants based on the definition presented by Agrawal et al.: ``\textit{Immersion is a phenomenon experienced by
an individual when they are in a state of deep
mental involvement in which their cognitive
processes (with or without sensory stimulation)
cause a shift in their attentional state such that
one may experience disassociation from the
awareness of the physical world}"\cite{agrawal2019defining}.  

\begin{figure}[ht!]
    \centering
    \includegraphics[width=1.0\linewidth]{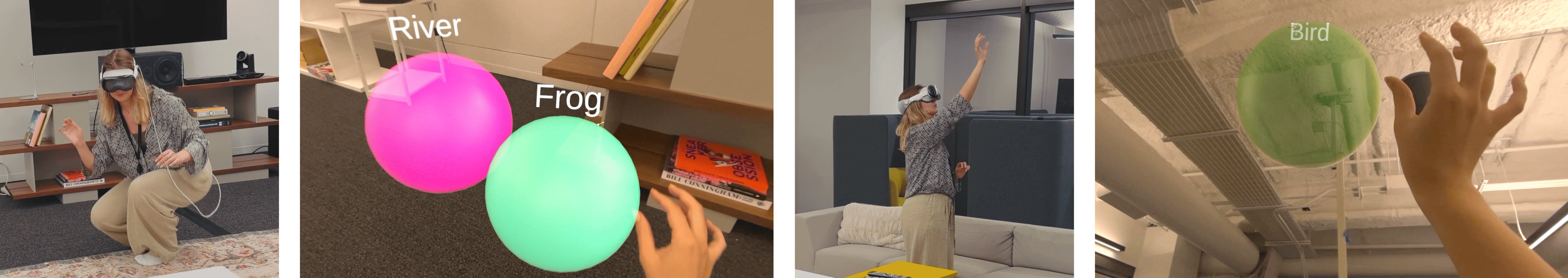}
    \caption{A user building a cinematic experience in the outdoor camping scene. Observe how the users crouches to place the \enquote{\emph{River}} and \enquote{\emph{Frog}} audio objects on the floor, and positions the \enquote{\emph{Bird}} audio object close to the ceiling.}
    \label{fig:environment-scene-example-1}
\end{figure}

In the \emph{cinematic scene} (Figure \ref{fig:environment-scene-example-1}), they positioned \enquote{\emph{Bird}} chirps, \enquote{\emph{Frog}} croaks, \enquote{\emph{River}} stream, \enquote{\emph{Breeze}}, and \enquote{\emph{Campfire}} sounds to simulate an outdoor camping environment. We randomized the order of these two scenes for each participant, and each scene had a 5-minute limit (or ended earlier if participants were satisfied of their mixing output). Throughout the sessions, participants were free to navigate the room, placing virtual objects anywhere in the physical room (from floor-level to near the ceiling) to get the sonic effect they envisioned. We recorded the positions of virtual audio objects and participant's head movements. External and internal HMD cameras captured a full account of each session for subsequent analysis. 

\subsubsection{Measures. } Upon completing each scene, participants filled out the NASA-TLX \cite{nasatlx1988} to report on mental, physical, and temporal demands, as well as perceived effort and frustration (refer to the full list of questions in appendix, \ref{sec:appendix}). We used a 7-point Likert scale, where 1-3 corresponded to lower workload, 4 was neutral, and 5-7 indicated higher workload. Participants also completed a \emph{demographic questionnaire} and a \emph{short-answer survey} (see appendix, \ref{sec:appendix}) describing any difficulties they found, suggestions for improvements, what they enjoyed the most, and -- if they were experienced DAW professionals -- potential applications of AudioMiXR in their own workflows. We also collected placement and movement logs (trajectories), identifying any patterns in how they arranged and organized virtual audio objects.

\subsection{Study 1 - Data Analysis Approach}
Two authors conducted a thematic analysis to identify themes in participant responses in order inform design lessons for mixing with 6DoF. Open coding was done independently and followed by axial coding to form larger code groups. Lastly, the two authors formed overarching themes specific to the interactive, aural, and visual elements with 6DoF. 

\subsection{Study 2 – 2D Panner Comparison}

To contextualize the usability, creative outcomes, and design insights from the initial study, we conducted a comparative experiment between AudioMiXR and a custom-built 2D panner interface developed in-house for desktop interaction.
\begin{figure}[H]
\centering 
\includegraphics[width=0.85\linewidth]{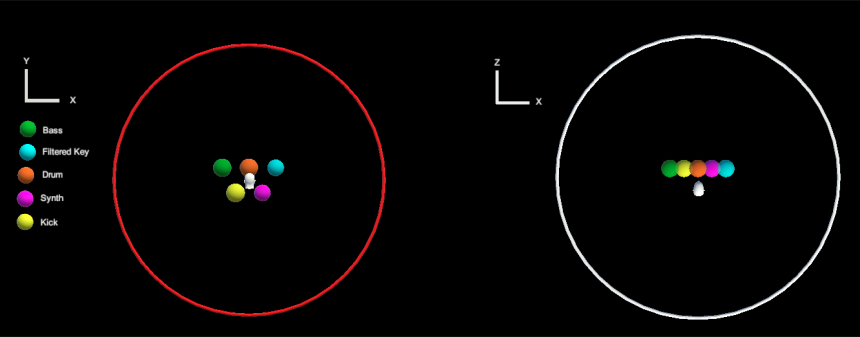}

\caption{\textbf{The 2D spatial audio panner. \textit{From left to right}}: \textit{Left: } Facilitates audio object panning in the vertical and horizontal axes and to the, \textit{Right: } audio objects can be panned in the horizontal and forward axes. This functionality parallels 2D desktop spatial audio panners present in Ableton Live, Reaper, FL Studio, and SPARTA DAWs. \cite{ableton2025},\cite{reaper2025},\cite{flstudio2025},\cite{sparta2025}}
\label{fig:2d_panner}
\end{figure}
\subsubsection{Setup. }
The experiment was conducted in a university lab environment without any physical obstructions that may interfere with the experiment. The 2D panner interface was displayed on a 27-inch ASUS ProArt monitor, with participants interacting via a standard wired mouse. Audio objects were represented as draggable spheres, analogous to those present in the AudioMiXR interface. The graphical user interface (GUI) of the 2D panner consisted of two circular planes that allowed users to manipulate the positions of audio objects along the $x$–$y$ and $x$–$z$ dimensions (see Figure \ref{fig:2d_panner}), following conventions commonly seen in 2D panners within widely used digital audio workstations (DAWs). Spatial audio objects are displaced in the same relative locations as the AudioMiXR interface with respect to the fixed origin. Spatial audio is rendered to the user depending on the directionality and distance of audio objects about the head. The same push-pull parameter control used in AudioMiXR is applied to the 2D panner, but instead uses a drag-and-drop function with a mouse. In AudioMiXR, users must push objects out to 4 m (i.e., 4 world units in Unity). We mirror this functionality in the 2D panner by implementing logarithmic distance-based attenuation, such that objects approaching the edge of the rings become quieter until they become silent. 

\subsubsection{Participants. }
We recruited 18 participants via snowball sampling from Brown University. The participant pool (N = 18; 9M, 9F) was composed of non-experts in audio production (mean age = 23.3 years, range = 18–37). 10 participants reported perfect vision, while 8 indicated they were nearsighted. This group was chosen to evaluate interface usability and creative potential in novice users without bias from prior tool familiarity.

\subsubsection{Procedure. }
Each participant received 3 minutes of training on each interface (order randomized), followed by 4 trials each 4 minutes long---composing both music and cinematic soundscapes in both AR and 2D (trial order randomized). The same audio stimuli were used in both interfaces. All sessions were conducted using AVP headsets, and spatialization was controlled via identical Unity spatial audio settings. Importantly, while the AR interface supports 6DoF interaction, we restricted it to 3DoF (rotation + orientation only) to match with the directional placement capabilities of the 2D interface---since the 2D interface does not allow for translation we designed the experiment to control confounding variables that may arise if the AR user can translate. Participants were instructed to create a soundscape that is most immersive to them based on the definition of immersion described by Agrawal et al. \cite{agrawal2019defining}.

\subsubsection{Measures. }
We assessed subjective workload and usability using two established instruments. Task-related workload was evaluated using the NASA-TLX as described in section, which includes five dimensions: mental demand, physical demand, temporal demand, effort, and perceived performance \cite{nasatlx1988}. Participants rated each NASA-TLX dimension on a 7-point Likert scale (1 = very low workload, 4 = neutral, 7 = very high workload), with higher values indicating greater task demand or lower perceived performance. The System Usability Scale (SUS) included ten items assessing perceived usability, also rated on a 7-point Likert scale from 1 (strongly disagree) to 7 (strongly agree) \cite{brooke1996sus}. Additionally, we assessed creative output by asking 7-point Likert scale questions informed by Study 1's results, probing participants’ perceived creativity, their inclination to explore spatial arrangements, and the extent to which visual representations of sound objects supported creative insight. Lastly, we asked a general user satisfaction question for both the AudioMiXR and 2D interface on another 7-point Likert scale (1 = very low, 4 = neutral, 7 = very high).

\paragraph{Data Analysis Approach}
For analysis, TLX, SUS, and creative output scores were rescaled to a standardized 0--100 range. Objective performance metrics included task completion time and the number of object edits. Because the study employed a within-subjects design, all comparisons were analyzed using paired non-parametric tests, with Holm corrections applied to control the family-wise error rate.

\section{RESULTS}

This section begins by presenting the quantitative and qualitative results of our exploratory AR study (Study 1), including NASA-TLX workload measures, statistical tests, and spatial visualizations of virtual audio object placement, which collectively offer objective insights about AR sound design patterns. We then present the qualitative findings from participant feedback, summarizing sentiments, codes, and themes on using virtual audio objects for sound design in AR across cinematic and music mixing tasks. To contextualize and validate these findings, we conducted a follow-up comparison study (Study 2) that quantitatively evaluates user experience across both the XR interface and a 2D desktop panner. Results from this complementary study provide a controlled contrast that highlights the specific usability and creativity benefits of immersive spatial interfaces.

\subsection{Study 1: Cognitive Demands}%

\begin{figure}[ht!]
    \centering
\includegraphics[width=1.0\linewidth]{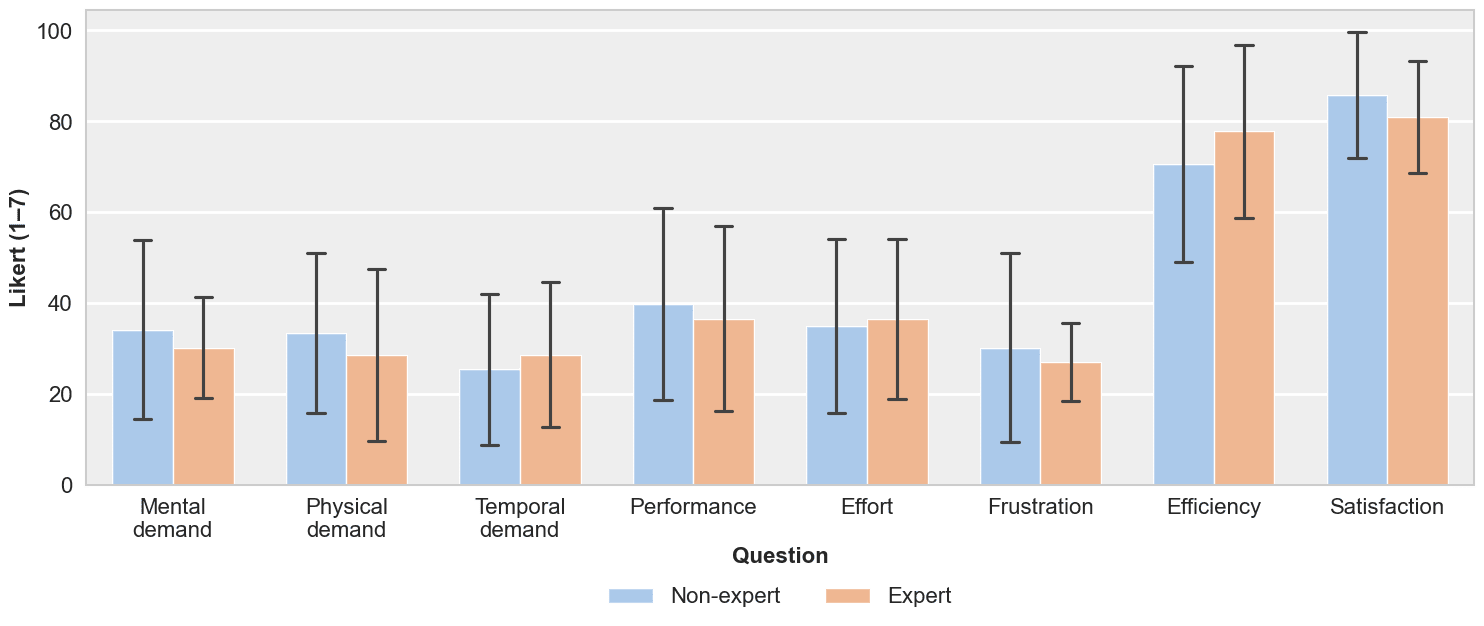}
    \caption{Comparison of mean Likert scores on NASA-TLX assessment profiled by DAW expertise level (with \emph{unexperienced}, \emph{beginners}, and \emph{intermediate level} users labeled as \enquote{non-experts}, and \emph{advanced} and \emph{expert} level users classified as \enquote{experts}).}
    \label{fig:nasa-tlx-experts}
\end{figure}

First we examined the normality assumptions with the Shapiro-Wilkinson test, which indicated that only the dimensions of Frustration, Efficiency, Satisfaction for DAW expert users followed a normal distribution; but the remaining dimensions and groups violated the normality assumption (all $p<0.05$). Consequently, we opted for the non-parametric Mann-Whitney U to assess the statistical significance of trends. A comparison of NASA-TLX scores between DAW expert and non-expert users, as depicted in Figure \ref{fig:nasa-tlx-experts}, revealed no statistically significant differences across all workload dimensions based on the Mann-Whitney U tests (all $p$-values $>0.05$), suggesting that AudioMiXR imposes a similar perceived workload on users irrespective of their prior audio mixing experience and is thus suitable even for novice users. 
Overall, participants reported low levels of mental demand ($\mu=2.30, \sigma=1.20$), physical demand ($\mu=2.22, \sigma=1.25$), and temporal demand ($\mu=1.85, \sigma=1.13$) when manipulating virtual audio objects in AR, along with low frustration levels ($\mu=2.04, \sigma=1.22$). These factors help explaining the high satisfaction scores ($\mu=5.89, \sigma=0.93$) with the system usability. Participants perceived a moderate level of effort ($\mu=2.48, \sigma=1.28$) required to complete the tasks, and the performance dimension received neutral ratings ($\mu=2.70, \sigma=1.44$), reflecting on the success of positioning and manipulating virtual audio objects accurately without excessive difficulty. Additionally, the ratings on the efficiency of completing the tasks are consistently high ($\mu=5.11, \sigma=1.45$), highlighting that participants were able to complete tasks without overhead. 


\subsection{Study 1: Coordinates Visualization}\label{heatmaps}

\begin{figure}[ht!]
\centering
\begin{subfigure}{0.49\linewidth}
\includegraphics[width=1.0\linewidth]{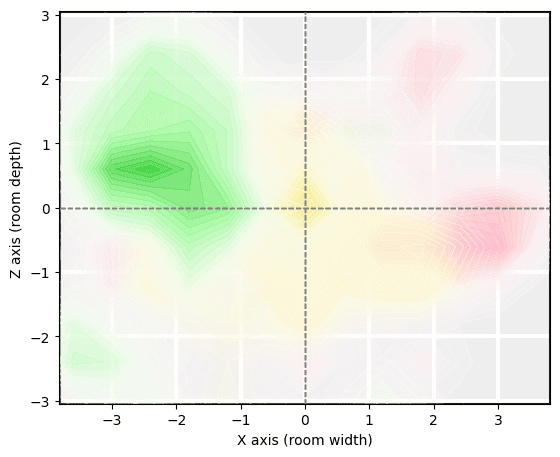}
\end{subfigure}
\hfill      
\begin{subfigure}{0.49\linewidth}
\includegraphics[width=1.0\linewidth]{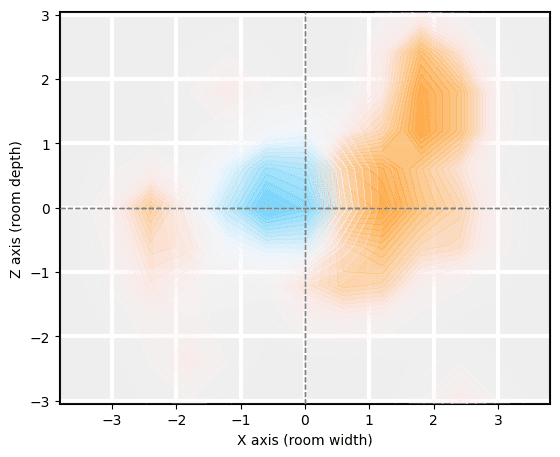}
\end{subfigure}
\\
\begin{subfigure}{0.49\linewidth}
\includegraphics[width=1.0\linewidth]{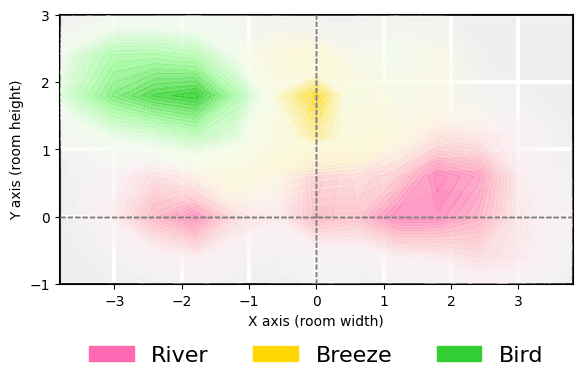}
\end{subfigure} 
\hfill
\begin{subfigure}{0.49\linewidth}
\includegraphics[width=1.0\linewidth]{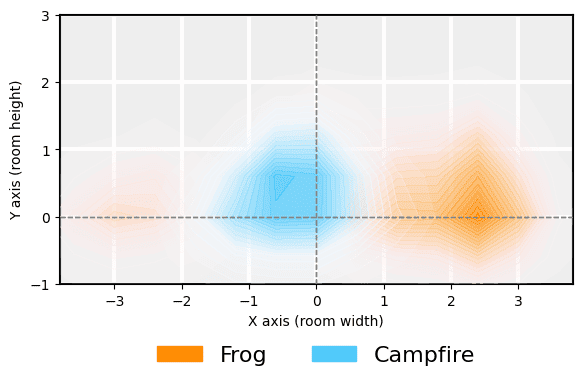}
\end{subfigure}
    \caption{Heat maps of the mean audio object 3D positions in the environment scene (first row: $X$-$Z$ plane, second row: $X$-$Y$ plane).}
    \label{fig:heatmap-environment-scene}
\end{figure}

\begin{figure}
\centering
\begin{subfigure}{0.32\linewidth}
\includegraphics[width=1.0\linewidth]{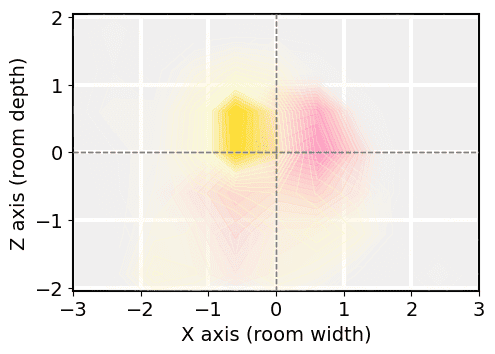}
\end{subfigure}
\hfill      
\begin{subfigure}{0.32\linewidth}
\includegraphics[width=1.0\linewidth]{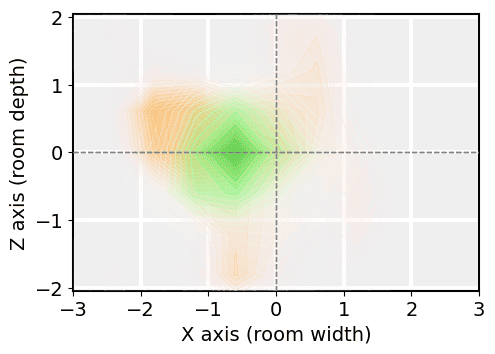}
\end{subfigure}
\hfill      
\begin{subfigure}{0.32\linewidth}
\includegraphics[width=1.0\linewidth]{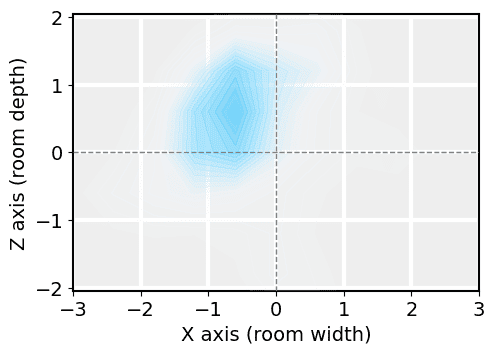}
\end{subfigure}
\\
\begin{subfigure}{0.32\linewidth}
\includegraphics[width=1.0\linewidth]{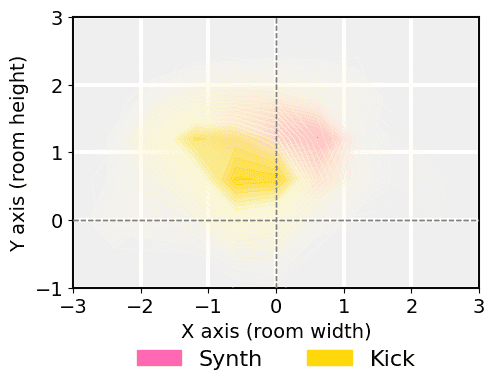}
\end{subfigure}
\hfill      
\begin{subfigure}{0.32\linewidth}
\includegraphics[width=1.0\linewidth]{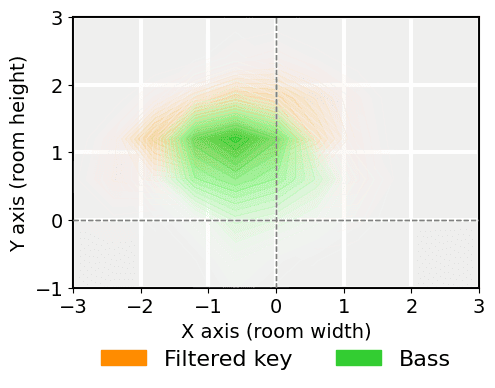}
\end{subfigure}
\hfill      
\begin{subfigure}{0.32\linewidth}
\includegraphics[width=1.0\linewidth]{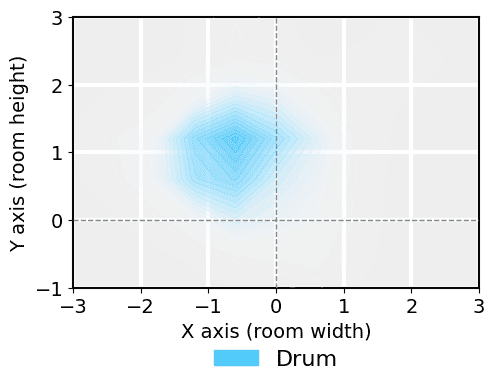}
\end{subfigure}
    \caption{Heat maps of the mean audio object 3D positions in the audio mixing scene (first row: X-Z plane, second row: X-Y plane).}
    \label{fig:heatmap-audio-mixing-scene}
\end{figure}

\begin{figure}
    \centering
    \includegraphics[width=1.0\linewidth]{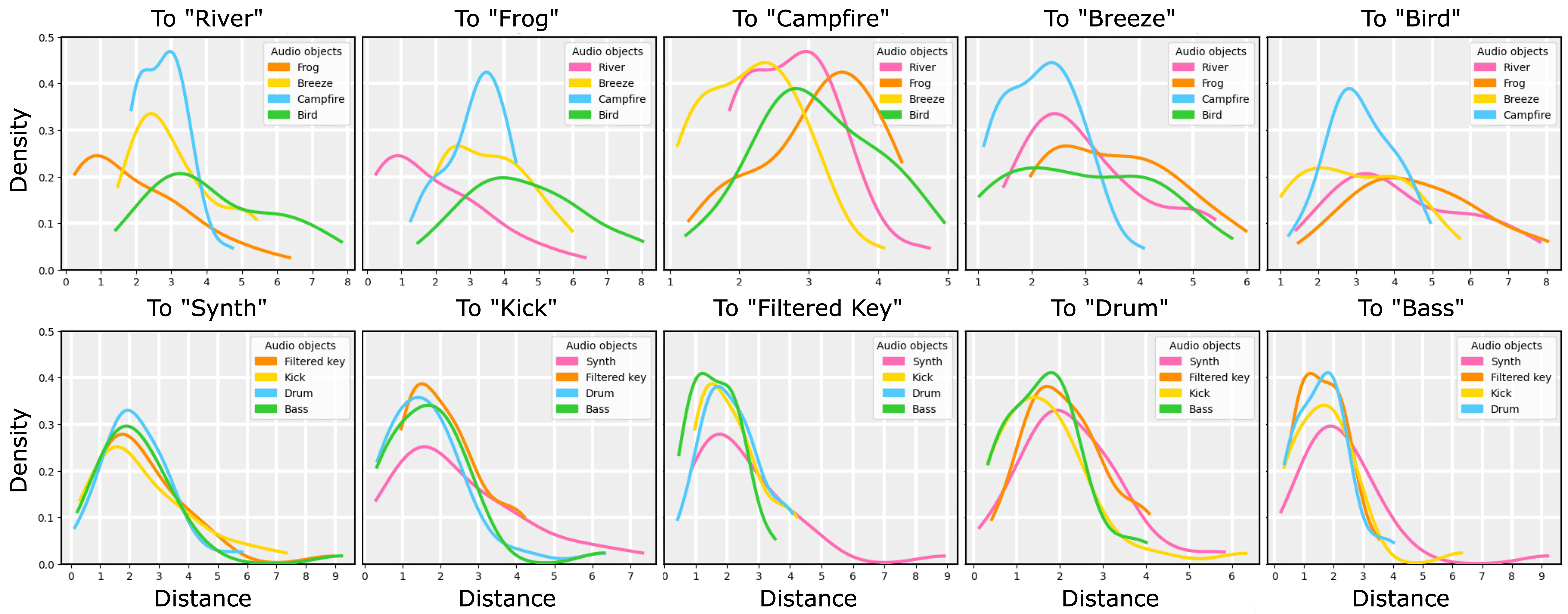}
    \caption{Density estimates using a Gaussian kernel over histograms of distances between pairs of virtual audio objects.}
    \label{fig:kdeplots}
\end{figure}

We analyzed the HMD logs from each user trial to aggregate the 3D coordinates of all virtual audio objects within each audio scene. This aggregation involved combining all logged timestamps across users and averaging the $x$,$y$,$z$ coordinates of each virtual audio object, thereby obtaining a mean 3D position for each object in each scene. Next, we binned the positions of each virtual audio object into two 2D histograms: one for the $X$-$Z$ plane (top-floor view) and another for the $X$-$Y$ plane (side view), and applied a Gaussian filter to the counts to produce smoother heat map visualizations of the average positions of each audio object class (Figures \ref{fig:heatmap-environment-scene} and \ref{fig:heatmap-audio-mixing-scene}). The color intensity within each heat map indicates the regions where a virtual audio object is more likely to be placed on average, across our trials. We observed significantly different spatial patterns between the two scenes, the cinematic and the music mixing scene, which are described below. We also computed the distances between pairs of virtual audio objects within each scene and estimated the densities of histograms of distances using a Gaussian kernel (Figure \ref{fig:kdeplots}).
These densities allowed us to compare the proximity relationships between pairs of virtual audio objects and correlate that with their semantics.

In examining the heat maps for the cinematic scene, as illustrated in Figure \ref{fig:heatmap-environment-scene}, the \enquote{\emph{Breeze}} sound is consistently positioned at the origin in $X$-$Z$ plane, corresponding to the location where the user initiates the application, and approximately at the headset height in the $X$-$Y$ plane. Because  \enquote{\emph{Breeze}} is an ambient sound, participants naturally tended to place it near their headset to emulate its pervasive presence. In contrast, the \enquote{\emph{Bird}} sound, characterized as a point-source sound, is generally positioned above the headset in the $X$-$Y$ plane. For instance, during the experiment a participant expressed they felt \emph{the birds should be placed up high like there are flying or in a tree}. The opposite occurs with the \enquote{\emph{River}} audio object, expected to emanate from the ground and, thus, heard from below. Consequently, in the $X$-$Y$ plane, it is typically situated on the floor. The \enquote{\emph{Frog}} audio object exhibits greater variance across both planes, likely because, as an animal, it could virtually be placed anywhere on the $X$-$Z$ and still be coherent with the scene. Nonetheless, it exhibited a considerable overlap with the \enquote{\emph{River}} heat map, indicating a tendency for participants to place the frog close to the river, reflecting nature-related biases. This is a possible manifestation of the Gestalt principle of similarity \cite{farnell2010designing, shelvock2016gestalt}, where the strongest proximity between subjects relates to those that have overlapping semantic associations, and led our participants to arrange those virtual audio objects together. This is also observed in the density estimates in Figure \ref{fig:kdeplots}, where the closest virtual audio object to \enquote{\emph{River}} is \enquote{\emph{Frog}}, and vice versa. Additionally, the \enquote{\emph{Frog}} audio object is generally placed below the headset height, similar to the \enquote{\emph{Campfire}} audio object. The \enquote{\emph{Campfire}} is typically positioned at the origin, close to the user. This pattern may stem from participants viewing the campfire as the central point of the camping setup, thus leaving it at the origin. Alternatively, it might be perceived as an ambient sound with a larger spread, justifying its proximity to the user.

In the music mixing scene (Figure \ref{fig:heatmap-audio-mixing-scene}), the heat maps present considerably more overlap and are primarily more concentrated around the origin (notice the range of the $X$-$Z$ axes), extending roughly to the user's reach envelope. 
In standard spatial audio panning the loudness of spatial audio objects is kept constant, only the room response is modified. 
In those scenarios, expert audio mixers \cite{gibson1997art, dewey2024} have noted that a common practice is to place high frequencies (called \enquote{treble} bands in sound design) higher and low frequencies (also known as \enquote{bass} bands) lower in the spatial field, reinforced by psychophysics research on congruence between pitch and visual vertical positioning \cite{bernstein1971, karla2010vision, crossmodal2016chi}.
However, we have not observed such a pattern with audio objects with stronger high-frequency content (e.g. the \enquote{\emph{Drum}}) versus audio objects with richer low-frequency content (such as the \enquote{\emph{Filtered Key}}), likely because loudness was not kept constant. In a 6DoF application, raising or lowering a virtual audio object spatially substantially increases its Euclidean distance from the headset, effectively causing larger loudness attenuation than if the object was kept at the HMD level. That might explain why we observe that most of the music mix audio objects are placed approximately at the HMD level, producing notable overlap among different objects (Figure \ref{fig:kdeplots}).

\subsection{Study 1: Variance Analysis} \label{variance-analysis}

\begin{figure}[ht!]
\centering
\begin{subfigure}{0.49\linewidth}
\includegraphics[width=1.0\linewidth]{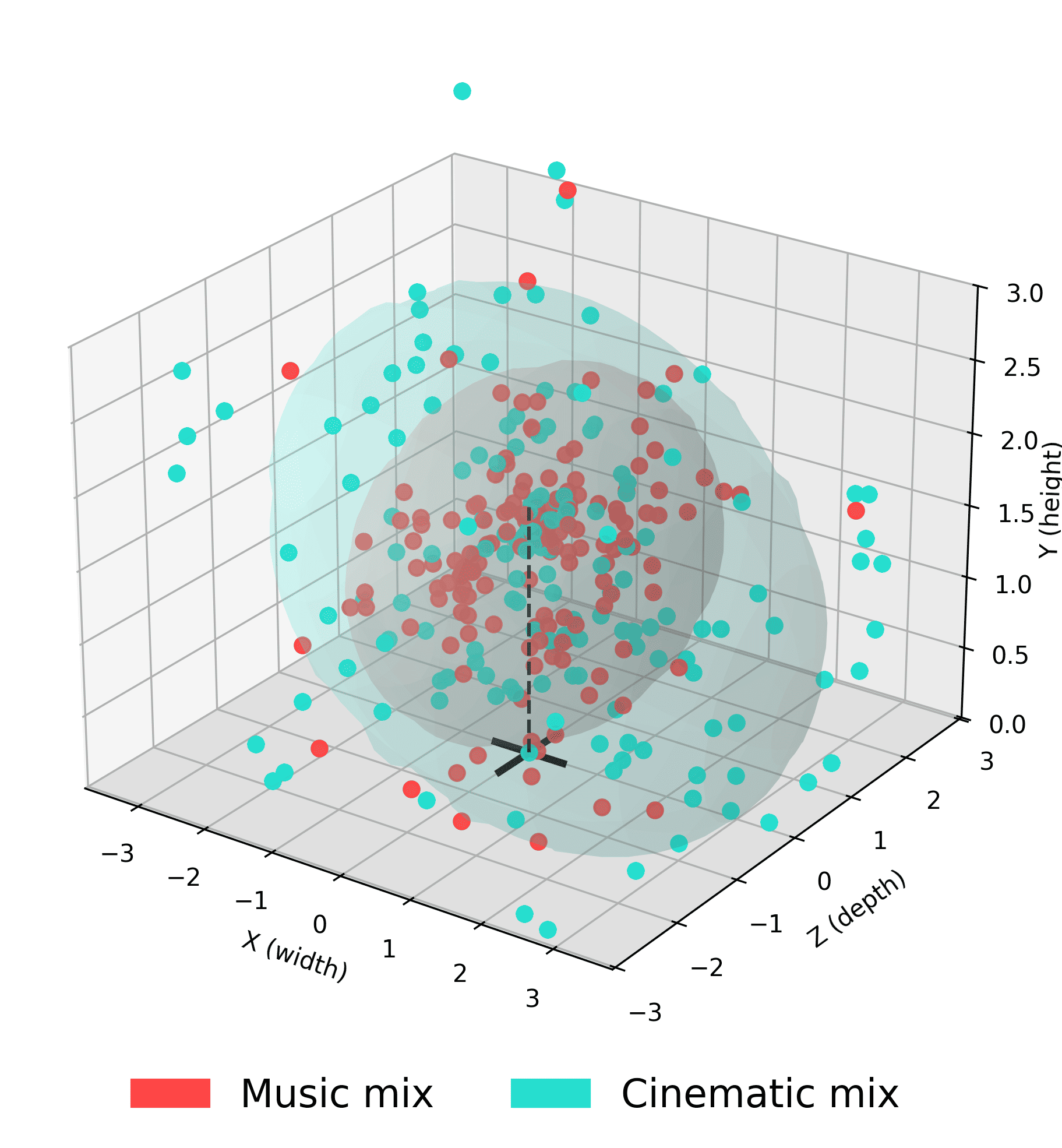}
\end{subfigure}
\hfill      
\begin{subfigure}{0.49\linewidth}
\includegraphics[width=1.0\linewidth]{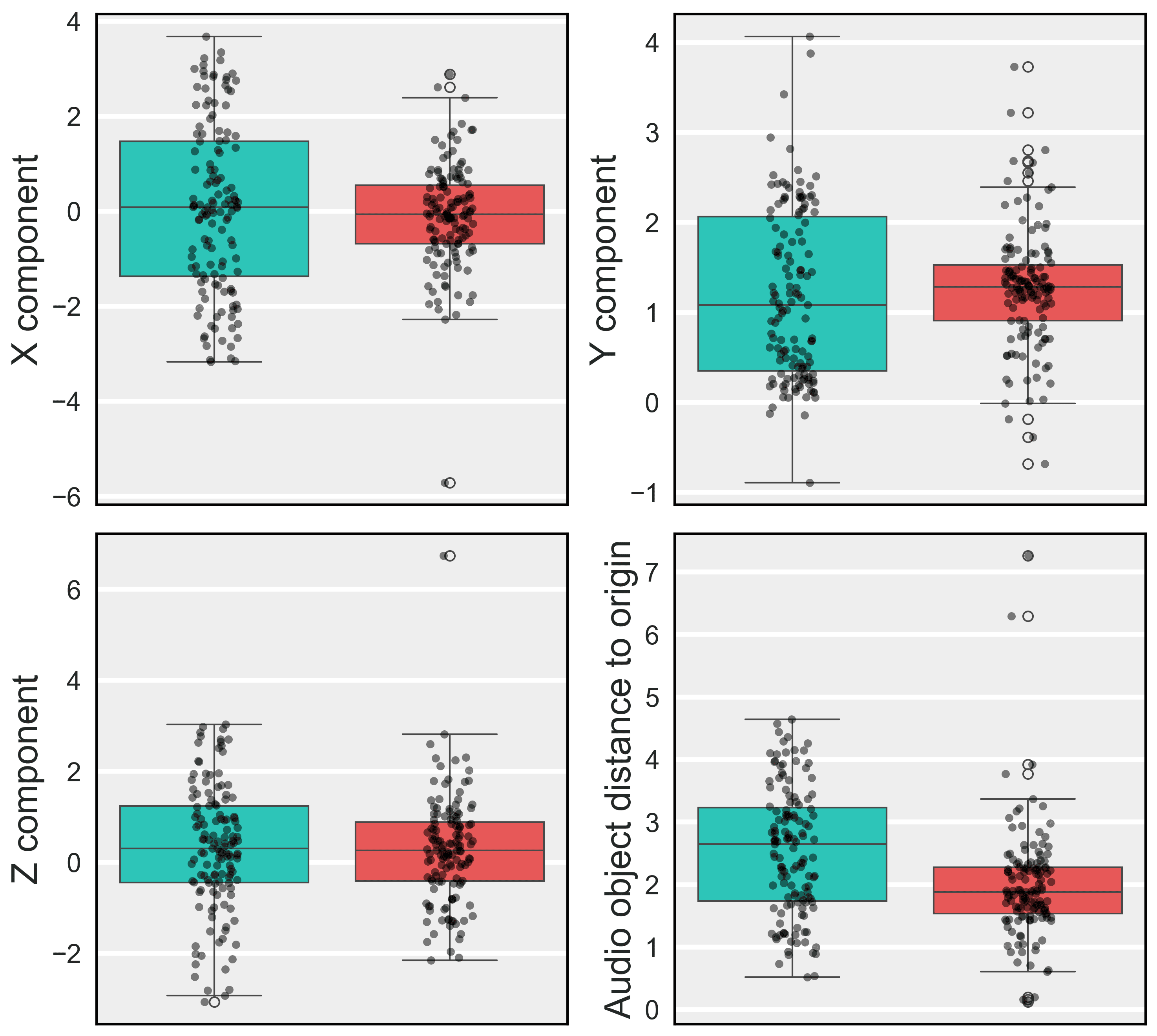}
\end{subfigure}
\caption{(Left) A room representation showing the final placement of audio objects across all experiments and users, color-coded by mix. A Gaussian mixture with two components (one for each mix) is fitted to the data. The cross at the origin represents the initial position, while the vertical dashed line represents the average user height; (Right) Boxplots illustrating the distribution of the final placement of audio objects along the $X$, $Y$, $Z$ axes, as well as the distribution of distances from the final audio object positions to the origin.}
\label{fig:tests-3d-gaussian}
\end{figure}

To assess the homoscedasticity of final audio object placement between music and cinematic mixes, 
we employed Levene's test. Specifically, we examined the distributions of $x$,$y$,$z$ spatial components of the final audio object placement, as well as the overall distances of audio objects from the origin (see Figure \ref{fig:tests-3d-gaussian}). 
Our analysis revealed $p$-values below the significance level of $\alpha=0.05$ for all tested components ($x$,$y$,$z$, and distance-based: $F_X=36.94$, $p<0.05$; $F_Y=30.52$, $p<0.05$; $F_Z=4.93$, $p<0.05$; $F_d=13.90$, $p<0.05$), indicating that the variances between the music and cinematic mixes are significantly different. 

To visualize and further understand these differences, we represented the experimental room setup with the final placements of virtual audio objects across all users and scenes in Figure \ref{fig:tests-3d-gaussian}. We fitted a multivariate mixture-of-Gaussian (MoG) model using the expectation-maximization algorithm for density estimation, choosing one component per mix to capture the distinct spatial distributions of music and cinematic mixes. Each component was assigned its own covariance matrix. The resulting 3D Gaussians ellipsoids (see Figure \ref{fig:tests-3d-gaussian}) illustrate the spatial dispersion of final audio object placements. The estimated means for the cinematic scenes ($\hat{\mu}_X=0.071$, $\hat{\mu}_Y=1.247$, $\hat{\mu}_Z=0.242$) and mix scenes ($\hat{\mu}_X=-0.064$, $\hat{\mu}_Y=1.324$, $\hat{\mu}_Z=0.209$) indicate slight positional shifts between the two mix types, while the variances ($\hat{\sigma}_X^2=2.654$, $\hat{\sigma}_Y^2=0.766$, $\hat{\sigma}_Z^2=1.610$ for cinematic and $\hat{\sigma}_X^2=1.016$, $\hat{\sigma}_Y^2=0.384$, $\hat{\sigma}_Z^2=1.156$ for music) reveal that audio object placement in cinematic mixes exhibit greater variability, particularly along the $X$ and $Z$ axes (width and depth dimensions of the room), corroborating the results from Levene's test. 

\subsection{Study 1: Qualitative Analysis}
This section outlines the findings from our thematic analysis, where we organized the codes into three themes: Embodied Mixing, Mapping Sound Attributes
to AR GUIs, and Prototyping with
Spatial Interface. These themes underscore the overarching insights gathered from participant responses to provide context for our analysis and design lessons. 
\begin{table}[ht!] 
    \caption{Final Codebook of Identified Themes, Corresponding Codes, and Exemplar Quotes}
    \centering
    \begin{tabular}{>{\raggedright}p{4cm}|>{\raggedright}p{4cm}|>{\raggedright\arraybackslash}p{6cm}}
        \toprule
        \textbf{Theme} & \textbf{Codes} & \textbf{Exemplar Quote} \\ \midrule
        \multirow{4}{4cm}{1. Embodied Mixing \textbf{(N = 20, 74\%)}} 
            & 6DoF Audio Presence \textbf{(N = 14, 52\%)} & 
                P19: \enquote{\emph{I liked being able to hear the sound coming from different directions as I moved it around, it is cool that instead of just hearing the sound around you you can interact with the sound }} \\ \cline{2-3}
            & 6DoF Audio Perceptibility \textbf{(N = 8, 30\%)} & 
                P6: \enquote{\emph{I chucked my head into a sphere, so I could hear just what was going on in there}} \\\cline{2-3}
            & Visual-Aural Responsiveness \textbf{(N = 7, 26\%)} & 
                P22: \enquote{\emph{There is a natural intuition
                on how to manipulate the objects. The immediate
                acoustic feedback made it easy to converge
                on manipulations that were most effective}}  \\ 
        \midrule
        \multirow{2}{4cm}{2. Mapping Sound Attributes to AR GUIs \textbf{(N = 18, 67\%)}} 
            & Audio Attribute Level Control \textbf{(N = 15, 56\%)} & 
                P24: \enquote{\emph{For
                example close distances always made the object
                much louder than I wanted. I’d like to specify
                volume across different distances more precisely}} \\ \cline{2-3}
            & Object Appearance \textbf{(N = 8, 30\%)} & 
                P23: \enquote{\emph{I think the size of audio object should be associated with spread}} \\ \cline{2-3}                & Modularity \textbf{(N = 4, 15\%)} & 
                P25: \enquote{\emph{ I could see this integrated as a plugin in my DAW while looping / experimenting with multitrack placement.}}\\ \midrule
        \multirow{2}{4cm}{3. Prototyping with Spatial Placement \textbf{(N = 12, 44\%)}} 
            & Vertical Positioning \textbf{(N = 7, 26\%)} & 
                P17: \enquote{\emph{ Make it easier to move objects far away. Make it clear to look up if user is trying to move an object overhead}} \\ \cline{2-3}
                   & Spatial Exploration \textbf{(N = 6, 22\%)} & 
                P25: \enquote{\emph{Also Experimented with dragging objects while moving – Moving sound around while walking throughout the space, also finding different mix levels within a room to compare by walking through it.}} \\ \cline{2-3}
                & Shaping soundscape \textbf{(N = 5, 19\%)} & 
                P6: \enquote{\emph{The wind should be everywhere, so how would I do I turn a wind to an object that isn’t spatial, it’s like a skybox in a computer game, so I put it above myself}} 
                \\ 
         \bottomrule
    \end{tabular}
\end{table}

\subsubsection{Theme 1: Embodied Mixing}

We first present codes describing participant responses (N = 20) related to their bodies' influence on their mixing experience. Overall, respondents expressed that the combination of their body movements and the feedback they received contributed to their experiential design process and supported their mix. 

\paragraph{6DoF Audio Presence}
About half of the participants (N=14; 8 Non-Expert, 6 Expert) shared positive sentiments highlighting the effects of 6DoF, which adapted to their position and orientation-induced feelings of \emph{presence} when designing their soundscape. We adopt the two-dimensional conceptualization of \emph{presence} suggested by Weber et al. and define it as, \enquote{as the allocation of
attentional resources to the mediated world and the sensation of perceptually being surrounded by the virtual environment or the user's judgment about the degree of realism} \cite{weber2021get}. Participants found the directional audio feedback to feel \enquote{\emph{real-time}} in response to their position and orientation. Additionally, some participants described enveloping themselves in their mix by navigating freely through the responsive soundscape or surrounding themselves with their audio objects: 
\begin{quote}
    \emph{\enquote{Being able to place the different parts of the environment where you wanted was super cool. In the camping environment, I closed my eyes and it felt like I was actually camping. Being able to put specific tracks/channels above and behind me in the music mixing environment felt like I was truly immersed in the song.} -- P21 (expert) }
\end{quote}

Participants noted that when they placed objects above or behind themselves, the 6DoF spatial audio accurately translated those physical positions into perceivable directional cues resulting in an \enquote{\emph{immersive}} experience. This aligns with research linking mediated experiences involving body movement in physical spaces to an enhanced sense of presence \cite{presence1997,shan2024work,woodard2025cam2camexploringdesignspace} -- specifically, when individuals engage more of their physical selves, the experience often feels more real. 
Some respondents delved deeper into why they were particularly intrigued by the sound's directionality, noting how it interacted with and changed based on their physical movements.: 
\begin{quote}
    \emph{\enquote{I liked being able to hear the sound coming from different directions as I moved it around, it is cool that instead of just hearing the sound around you can interact with the sound.} -- P19 (Non-Expert)}
\end{quote}
The interface may have added a level of interactivity for experts and non-experts, which appeared to be unique from their everyday audio experiences, possibly contributing to feelings of envelopment. The dynamic audio changing due to their movements can be attributed to how the spatial audio responds to their body with 6DoF, which allows for more precise directionality and includes translation (e.g., audio attributes changing as a user walks away from origin) versus 3DoF interfaces which, in the context of audio mixing applications, generally support only rotational components.

\paragraph{6DoF Audio Perceptibility} 
Some participants (N = 8; 5 Non-Expert, 3 Expert) shared 
specific strategies leveraging spatial audio with 6DoF, free-hand manipulation, and head orientation enabling strategies of perceving and isolating sound emitted from audio objects. Generally, increased users' perceptibility was due to the physical actions coupled with binaural rendering, amplifying interaural level differences (ILD). ILD is a phenomenon where the head occludes acoustics depending on the direction, altering the amplitudes' distribution to either ear \cite{farnell2010designing,birchfield2005acoustic}. For instance, during the experiments, people often moved the relative position between their ears and the audio objects to hear different perspectives of their mix: 
\begin{quote}
    \emph{\enquote{I really enjoyed the head tracking aspect. I felt that being able to bring an object close and then tilt my head from side to side made it easier to get a sense of the timbre of the sound. Also, being able to isolate objects in space was helpful.} -- P22 (Non-Expert)} 
\end{quote}
Head orientation plays a key role in sound localization, and actions like turning the head toward or away from the source assist in determining its physical location\cite{farnell2010designing,birchfield2005acoustic}. Therefore, head movement may have felt advantageous to users when assessing the quality of their mix.
Another technique worth noting was expressed by an expert during the experiment where they state,
\begin{quote}
    \emph{\enquote{I tried putting my head inside the audio sphere to isolate the sound.} -- P21 (Expert)}
\end{quote} 
P21 defined their strategy as inserting their head inside a virtual audio object, thereby attenuating all other sound sources within the scene. The sound isolation likely happened because the audio listener (attached to the AVP's position) was dominated by the audio source of the object where P21 inserted their head due to the distance attenuation in Unity's 3D sound settings. In the sound settings, other audio sources attenuate according to a logarithmic distance scale. As a result, sources that are in close proximity, or in this case near the same position, will dominate the audio output. All user responses in the coded group shared similar statements about the ease of perceptibility, except for one, who noted that the stark differences in audio could cause some confusion when mixing:  
\begin{quote}
    \emph{\enquote{ I found some inconsistencies with how I wanted the scene to sound in my head when looking at the objects versus how the sound actually rendered out. Moving around an object (walking in a circle around a ball) made me question the placement of other objects in relation to it because the changes in level were semi-stark.} -- P27} 
\end{quote}
P27 describes the disorientation they experienced due to conflicting expectations of where they observed their placement versus the sound they heard. The misalignment was exacerbated by walking in a circle, which likely caused the 6DoF audio to pan too sensitively in the user's headphones, hindering their ability to perceive the audio sources coherently and thus affecting their expected mix. To address this sentiment, it may be worth offering users greater flexibility over the sensitivity of the spatial audio panning.

\paragraph{Visual-Aural Responsiveness} 
Another subset of participants (N = 7; 4 Non-Expert, 3 Expert) found the alignment of the position of the audio objects' visual appearance corresponding to changes in the audio directionality, specifically when they would manipulate objects with free-hand gestures, to be both useful and enjoyable. Users remarked the immediate audio and visual feedback in response to their free-hand manipulation as \emph{intuitive} and assisted them in mixing, allowing for easy self-correction of placements:
\begin{quote}
\emph{\enquote{I really enjoyed the lack of latency when moving objects around the space. It really felt like the objects responded to my gestures in real-time, and their movements were also accurately reflected by the audio.} -- P23}
\end{quote}
P23's response explicitly highlights the sense of control in feeling convinced the audio objects reacted visually and aurally to their physical manipulations within a time window, making the physical-virtual interaction believable. Congruent audio-visual feedback has been shown to result in better performance and higher engagement \cite{crossmodal2016chi, o2024sound}. The blending of senses, also known as \enquote{synesthesia}, has been previously linked to excitability and arousal \cite{Terhune2011}, a fact that film, games and VR entertainment creators have been exploiting -- by integrating cross-modal stimuli, they evoke a sensation of presence that creates exciting experiences \cite{presence1997, o2024sound}, and if spent an extended period of time in the immersive experience, virtual artifacts start feeling like physical \cite{finnegan2016distance}. One of our participants expressed a synesthetic experience: 
\begin{quote}
\emph{\enquote{I quite like how close I can be to the audio objects (it almost feels like there's a haptic feedback from being close to certain audio objects)} -- P16}
\end{quote}
Another participant echoes this sentiment and also expressed how the immediate visual-aural feedback assisted their mixing process: 
\begin{quote}
\emph{\enquote{The immediate acoustic feedback made it easy to converge on manipulations that were most effective.} -- P22 (Expert)} 
\end{quote}
Overall, the immediate visual-aural feedback and alignment were complimentary in providing interactions that allow users to feel in control, thereby making their mixing process feel like a \emph{direct} result of their intended placements. The interrelationship of the visual and aural attributes of the interface captured in participant responses highlights the need for proper alignment and immediacy of feedback; otherwise, the mixing experience may be disrupted. Additionally, multimodal synchronization is a key sound design principle necessary to reduce cognitive load when, for example, viewing and listening to cinematic experiences or when developing audio authoring tools \cite{farnell2010designing}.

\subsubsection{Theme 2: Mapping Sound to AR GUIs}
Next, we outline the codes centered on controlling sound attributes in an AR interface. Responses (N = 18; 10 Non-Expert, 8 Expert) in this category generally consisted of the trade-offs of using depth to modify audio attributes (e.g., loudness or reverberation), instead of varying alternative visual properties and physical interactions of the audio objects to have better UI affordances and enable more granular control.

\paragraph{Audio Attribute Level Control}
The \emph{Audio Attribute Control} centers on challenges that arose when using the \emph{push-pull} action to vary the loudness parameter coupled with suggestions to modify alternative audio attributes to enable more fine-tuned control over their soundscapes. In this category, users (N = 15; 9 Non-Expert, 6 Expert) noted difficulties in depth perception when pushing objects out to control the loudness parameter of the audio objects:
\begin{quote}
\emph{\enquote{Depending on where I was standing in the room, I wished there was some guidance on how far I could push out an object on each $XYZ$ plane before they became inaudible/not part of the mix. } -- P27} 
\end{quote}
P27 highlights an issue with recognizing when an audio object becomes \enquote{inaudible}. This sentiment is aligned with how multimodal cues are often explored for additional guidance through an interface, and how depth alone may not offer enough control over audio attributes for some users \cite{chen2018investigating,hashky2024multi}. In this particular case, the user expresses uncertainty when audio objects will become inaudible, and their response suggests the inclusion of additional visual cues, like implementing a visual boundary representing the \emph{inaudible zone} that persists through the duration of the application to identify the distance before an object is silenced quickly. 


Participant responses in this category also expressed mostly negative sentiments regarding the \enquote{loudness} parameter control implemented in the interface. Users generally felt pushing objects \emph{too far away} made it harder to align their hands with the perceivably smaller objects due to the camera's perspective view in Unity. P2 describes the difficulty of aligning their hands with the virtual audio object that has been pushed far away from them:
\begin{quote}
\emph{\enquote{At one point, I had moved some objects really far away from me, and I was struggling to bring them back. I had to physically walk over to the objects to bring them back.} -- P2 (Non-Expert)}.     
\end{quote} 

Moreover, determining the sensitivity using the forward axis for loudness combined with the challenges in perspective projection presented difficulties in varying the audio attribute. P16 states:
\begin{quote}
\emph{\enquote{It would be nice to have a way to tweak the sensitivity concerning distance. It took some trial and error to figure out how far I needed to move the object to get the level of loudness I wanted.} -- P16}
\end{quote}
Other participants echoed this sentiment and typically suggested different features like modifying the sensitivity of the \enquote{\emph{push-pull motion}} or alternative methods for level control that do not rely on distance.

\paragraph{Object Appearance}
In contrast to the \emph{Audio Attribute Level Control}, the \emph{Audio Object Appearance} categorizes sentiments related to participants' (N = 8; 4 Non-Expert, 4 Expert) preconceived associations with mentally mapping visual elements to audio attributes (e.g., loudness, reverb, or spread). Most of the respondents in this category described preferences of \emph{changing the size of the sphere to increase/decrease loudness} or other audio attributes not implemented in the interface. Additionally, some responses related to the trade-offs of using translucence to represent the loudness:
\begin{quote}
    \emph{\enquote{I think the size of the audio object should be associated with spread} -- P27 (Expert)}
\end{quote}

Their mixing experience with DAWs may influence P27's sentiment because the spread is typically associated with the size of audio objects in object-based spatial audio metadata, such as Dolby Atmos \cite{dolby2018atmosrenderer}. In contrast, a response from P22 (Expert) associated audio object size with \emph{draping} and felt translucence could represent \emph{direct-to-reverberant ratio}. The other two experts in this category also stated that size and translucence make sense when representing loudness. While there were mixed sentiments on the preferred visual elements for audio objects, all participants in this category suggested varying visual associations with different audio attributes, indicating that there may be associations from their mixing experience or, in the case of non-experts, there are natural associations with visual properties of the spheres and audio attributes:

\begin{quote}
\emph{\enquote{I wasn't sure if I could change the magnitude of the \enquote{orb} or if I need to bring it closer to increase the volume} -- P15 (Non-Expert)}    
\end{quote}
 P15 expresses confusion regarding the visible size of the \enquote{orb} being associated with loudness. The visual parameter, \emph{size} being mistakenly associated with \emph{volume} or loudness was also expressed by P23 and it may be due to the natural associations humans have between audio and visual representations, often referred to as audiovisual perception \cite{malpica2020crossmodal, crossmodal2016chi, karla2010vision, gibson1997art}. This could also contribute to a GUI design strategy focused on how to represent sound visually in the most effective way for the intended task. 

\paragraph{Modularity}
 The AudioMiXR interface was seen by four experts (N = 8; 4 Non-Expert, 4 Expert) as a component for a potential hybrid system combined with an existing DAW for audio mixing with more spatial awareness. The four experts expressed that designing within the execution environment could fit nicely in-game audio design or XR workflows and foresee the \emph{AudioMiXR} interface integrated into a hybrid system where they have the flexibility to switch between their DAWs on a desktop and into a headset when they require more spatial awareness.  For example, P27 states: 
 \begin{quote}
 \emph{
 \enquote{Think this fits nicely within game audio design workflows and specific audio mixing workflows where the output is specifically for AR/XR especially at the ideation phase of the project... DAWs are notoriously cumbersome to use unless you are experienced with hot keys and shortcuts so this was very refreshing to be able to make large changes with little physical exertion.}
 -- 27 (Expert)}
 \end{quote}
 Respondents thought of \emph{AudioMiXR} as a complement to their existing workflows where they can overcome some of the drawbacks of DAWs but felt they would still want access to their traditional mixing tools. The flexibility of control underscores an optimal medium where all audio attributes may not necessarily need to be mapped to the AR interface and instead can leverage its features when required.  

\subsubsection{Theme 3: Prototyping with Spatial Placement}
Lastly, we present responses related to the potential affordances and drawbacks spatially placing objects has on prototyping mixes. Participants (N = 12; 5 Non-Expert, 7 Expert) expressed suggestions, challenges, and benefits the existing interface has on prototyping their soundscapes. 

 \paragraph{Vertical Positioning}
\textit{Vertical positioning} refers to sentiments centered on audio object placement along the vertical axis affected participants' (N = 7; 5 Non-Expert, 2 Expert) (i.e., placing an object above or below their line of sight). P5 and P20 both expressed the features was \emph{playing with height} while the rest of the respondents voiced difficulties with placement along the vertical axes: 
\begin{quote}
\emph{\enquote{Placing object overhead was a bit awkward with the limited field of view. You can also feel the mass of the headset when doing that} --  P9 (Non-Expert) }   
\end{quote}
Placement of objects overhead or below the HMD is a unique feature supported by AudioMiXR. Still, the limited FOV forced users to engage in excessive upward head rotation so that the cameras on the headset could capture their hand gestures, thereby disrupting their placements. The difficulties with the vertical placement due to the FOV of the headset may imply that there should be alternatives to object placement. For example, a function that maps subtle movements along the vertical axis may be a potential circumvention to take advantage of the vertical placement unique to an audio mixing interface with 6DoF.

\paragraph{Spatial Exploration}

Respondents expressed how the autonomy and \emph{freedom} provided by the AudioMiXR interface encouraged exploration and experimentation of their arrangements while designing their mix (N = 6; 4 Non-Expert, 2 Expert). Responses center on free-hand movement and being able to use \emph{all} of the surrounding space to manipulate objects: 
\begin{quote}
    \emph{\enquote{It allowed me to be creative and place the objects as far or close as I wanted. I felt that it understood what I wanted it to do most of the times.} -- P8 (Non-Expert) }
\end{quote}

P8's sentiment represents the code group in expressing how the interface feels unrestricted to their movements. P12 describes how the spatial placement enabled them to \emph{remix} audio objects easily and try different combinations of loudness and spatial configurations, thus supporting their creativity. As metaphorized in \cite{shelvock2016gestalt}, a mix is a \enquote{sonic portrait} and its essence can be conveyed in a myriad of ways. This flexibility may be worth considering for creative workflows within AR Sound Design. Another interesting connection is related to the quantitative results on virtual audio object spread reported in Subsection \ref{variance-analysis}, where a significant difference in spread and, thus, spatial exploration between the cinematic and music mix scenes can be observed, which underscores that the amount of spatial exploration is task-dependent. 

\paragraph{Shaping of Soundscape}
During the experiment, a few participants (N = 5; 3 Non-Expert, 2 Expert) described different strategies of grouping or arranging multiple audio objects, as well as how these objects worked together to create a cohesive soundscape. For instance, P2 would have preferred the ability to \enquote{tie objects} to their location instead of the \enquote{physical environment} in order for the arrangement of objects in the scene to be relative to the headset which is more aligned with an audio experience with 3DoF. Further, the rest of the participants echoed different types of arrangement functionalities centered on different types of grouping of objects to make the design of their soundscapes more flexible. P27 captures these sentiments well by suggesting moving objects together to simplify their mixing process: 
\begin{quote}
\emph{\enquote{Grouping objects -- link multiple balls together and moving as a unit.} -- P27 (Expert)}
\end{quote}
P27 describes a feature that may embody the concept of a cohesive soundscape better than simply treating each audio object as separate entities at all times. It is possible P27's experience with DAWs inspired their suggestion as modern DAWs like Logic Pro support users to arrange audio channel groups \cite{apple2024logicpro}, which is a known technique for professional audio mixers to design complex soundscapes \cite{farnell2010designing, gibson1997art}. This would exhibit the Gestalt principles of proximity and common fate \cite{farnell2010designing}, where similar sounds -- where similarity may be measured in terms of rhythm, timbre, or semantics -- would have the tendency to be arranged and moved together. This is something that has been reflected in our quantitative data (Subsection \ref{heatmaps}) as well, where one can clearly see several instances of semantically-related audio objects to be clustered together in space.

\subsection{Study 2: Quantitative Results}

We analyzed quantitative differences between the XR interface and the 2D panner using paired-sample t-tests with Holm-corrected $p$-values to account for multiple comparisons. Measures included perceived workload (NASA-TLX), system usability (SUS), and creativity.

\subsubsection{NASA-TLX}

Participants reported significantly lower cognitive workload when using the XR interface compared to the 2D panner. Specifically, XR showed significantly lower scores in frustration ($t(11) = -3.73$, $p_\text{holm} = 0.0115$, $d = -0.88$), mental demand ($t(11) = -3.19$, $p_\text{holm} = 0.0278$, $d = -0.75$), and effort ($t(11) = -2.47$, $p_\text{holm} = 0.0576$, $d = -0.58$). The composite workload score also favored XR ($t(11) = -2.95$, $p_\text{holm} = 0.0362$, $d = -0.69$). While performance and temporal demand also trended toward improvement, these differences did not reach statistical significance under Holm correction. These results suggest that the XR interface imposes less cognitive load for spatial audio tasks than a 2D panner.

\begin{figure}

  \centering
  \includegraphics[width=0.9\linewidth]{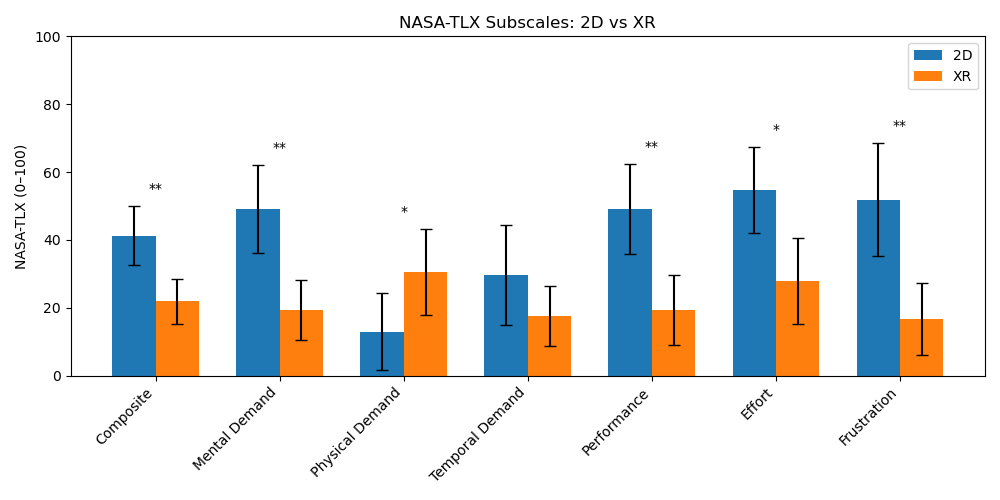}
  \caption{NASA-TLX workload comparison. XR showed lower cognitive load across key subscales. \textbf{Note: } ``Composite" refers to the overall distribution across all NASA-TLX categories.}
  \label{fig:nasatlx}
\end{figure}
\subsubsection{System Usability Scale (SUS)}

The XR interface was rated as significantly more usable across several SUS dimensions. Participants gave higher ratings for overall usability ($t(11) = 5.31$, $p_\text{holm} = 0.0005$, $d = 1.25$), confidence in use ($t(11) = 4.78$, $p_\text{holm} = 0.0014$, $d = 1.13$), learnability ($t(11) = 5.48$, $p_\text{holm} = 0.0004$, $d = 1.29$), and ease of use ($t(11) = 4.43$, $p_\text{holm} = 0.0023$, $d = 1.04$). The XR interface was also seen as significantly less cumbersome ($t(11) = -4.49$, $p_\text{holm} = 0.0023$, $d = -1.06$), and more conducive to frequent use ($t(11) = 6.99$, $p_\text{holm} < 0.0001$, $d = 1.65$). While some dimensions such as “need for support” and “getting started” did not differ significantly, the large effect sizes observed for many subscales indicate strong usability advantages for XR.

\begin{figure}

  \centering
  \includegraphics[width=0.85\linewidth]{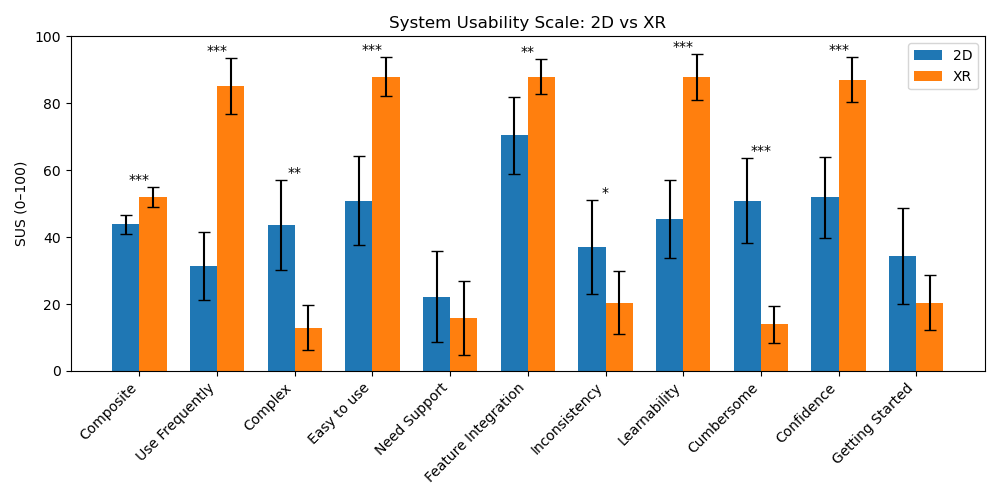}

  \caption{SUS scale results comparing 2D and XR interfaces (Holm-corrected $p$-values). \textbf{Note: } ``Composite" refers to the overall distribution across all SUS categories.}
  \label{fig:sus}
\end{figure}
\subsubsection{Creative Output}
\begin{figure}[H]

  \centering
  \includegraphics[width=0.9\linewidth]{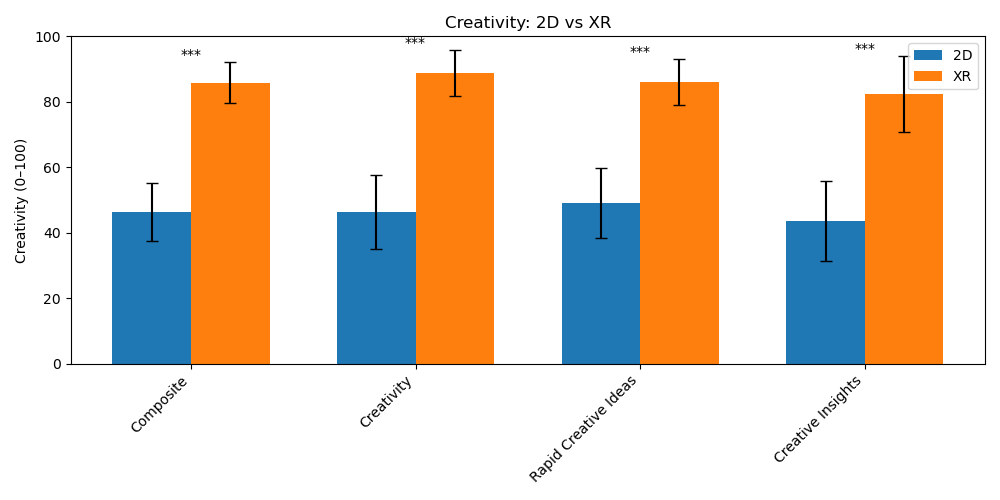}
  
  \caption{Creativity scale results. XR outperformed 2D on all creative subdimensions (*** $p < 0.001$). \textbf{Note: } ``Composite" refers to the overall distribution across all creativity categories.}
  \label{fig:creativity}
\end{figure}
The XR interface significantly outperformed the 2D panner across all creativity measures. Participants reported higher overall creativity scores with XR ($t(11) = 6.06$, $p_\text{holm} < 0.0001$, $d = 1.43$), and reported generating more rapid creative ideas ($t(11) = 5.33$, $p_\text{holm} = 0.0001$, $d = 1.26$). They also attributed more creative insight to the XR experience ($t(11) = 6.02$, $p_\text{holm} < 0.0001$, $d = 1.42$). The composite creativity score exhibited a large effect size favoring XR ($t(11) = 7.69$, $p_\text{holm} < 0.0001$, $d = 1.81$). These results suggest that the immersive and visual affordances of XR interfaces may substantially enhance creative engagement during spatial audio tasks.

\section{DISCUSSION}
In this section, we form design lessons derived by combining themes based on participant responses, analyzing our quantitative data, and sharing observations during the experiment while supporting these design lessons with existing literature in audio augmentations in AR, and authoring in XR. To complement these insights, we introduce a final subsection that presents findings from Study 2, a follow-up comparison study evaluating the 6DoF AR interface against a 2D panner baseline. This quantitative comparison strengthens our claims by highlighting where AR offers concrete advantages in usability, cognitive load, and creative potential. Afterward, we describe the types of mixing applications that may benefit most from a 6DoF AR interface, informed by both qualitative and quantitative findings.

\subsection{Design lesson -- Proprioception for AR Sound Design} 

Most users shared positive sentiments related to their experience when \enquote{\textit{freely walking around through their soundscapes}}, \enquote{\textit{hearing spatial audio from all directions change}} according to their movements, \enquote{\textit{immediate audio feedback}} in response to their actions and changes in the audio object positions, and \enquote{\textit{being able to explore and be creative}} which may inform how sound design principles should be translated into an AR setup. Theme \#1 highlights how users' bodies influencing the mix made them feel actively engaged in the directed tasks where they noticed their movements and actions directly affected the soundscapes they designed. Their fascination with their body's influence on the soundscape was primarily attributed to immediate UI response to participants' actions, position, and orientation, which are inherent to the 6DoF interactions in the interface and can facilitate sound design principles like sound attribute perception, layering and spatial awareness \cite{farnell2010designing,gibson1997art}. For instance, the dynamic visual-aural feedback in response to participants' movements seemed to allow for feelings of self-agency and control of the UI throughout the mixing tasks, as expressed in P23's statement: \enquote{\textit{It really felt like the objects responded to my gestures in real-time, and their movements were also accurately reflected by the audio}} in regards the immediate feedback in response to their physical actions.  Similar reactions align well with existing literature on perceiving ILDs and assist with audio localization, which can support the spatial audio mixing process \cite{birchfield2005acoustic,farnell2010designing,francart2007perception}. A 6DoF interface may be able to leverage proprioception in assisting the user in understanding the spatial relationships between their body, actions, audio objects, and the physical space around them.

Moreover, designing dynamic sound design interfaces in AR requires a use case that effectively leverages the space of the virtual environment. In Figure \ref{fig:tests-3d-gaussian}, we demonstrate the spatial distribution of all audio objects across all participants in the music mix and cinematic mix tasks -- participants generally used significantly more of the physical space to place audio objects for the cinematic blend. Additionally, we have provided a visualization of heat maps (Figures \ref{fig:heatmap-environment-scene} and \ref{fig:heatmap-audio-mixing-scene}) demonstrating where participants typically placed audio objects in the cinematic scene and uncovered that their placements aligned with real-world expectations of where participants could imagine the audio source (e.g., bird) would be concerning their physical location. The quantitative data demonstrates a possible benefit of using an AR design interface for mixing tasks that have familiar contexts in the real world, suggesting that this may be a reasonable focus for AR sound design UIs with 6DoF interactions. Further, our quantitative findings align well with some of our expert responses where they suggest using the AudioMiXR interface for soundscapes intended for \textit{experimenting with object positions in VR experiences} and \textit{video game design workflows}, which both use cases that involve a `perceivably' physical virtual environment where 3D position and orientation of audio objects would be necessary. 

\subsection{Design Lesson: Balancing Audio-Visual Modalities in AR GUI}
A number of feature requests emerged from our user study that reflect established psychophysical principles related to spatial cues, crossmodal congruences, and audiovisual localization -- all of which play a significant role in how participants want to experience sound design in AR. By addressing these features, AudioMiXR can better align with human perception, ultimately enhancing presence, immersion.

Several participants expressed alternatives for parameter control that align with familiar associations between visual effects and audio. Participants desired to control volume by \enquote{\textit{making objects larger or smaller}} or the \enquote{\textit{level of translucence}} of the audio objects, which are common associations present in traditional DAWs or music production tools \cite{gibson1997art}. Further, associations of visual and audio attributes are commonly researched in audiovisual perception \cite{crossmodal2016chi}. They may be coupled with AR design frameworks to design AR sound design interfaces sensitive to humans' preconceived associations with audio attributes and their visual representations. 

Participants consistently expressed the desire to have independent control of an audio object's loudness, as opposed to just moving it closer or farther away, 
as reflected by the following responses of P24: \enquote{\emph{I'd like to have control over object size and overall object volume at different distances [...] I'd like to specify volume across different distances more precisely}} or P27 \enquote{\emph{Expanding and reducing objects to represent volume change rather than just using distance to control levels}}. 
reflecting the strong cross-modal tendency to associate larger visuals with louder sound and viceversa \cite{farnell2010designing,gibson1997art}. Users also noted the challenge of distance-loudness calibration, (P24 \enquote{\emph{For example close distances always made the object much louder than I wanted}} or P16 \enquote{\emph{It would be nice to have a way to tweak the sensitivity with respect to distance. It took some trail and error to figure out how far I actually need to move the object to get the level of loudness I want}})
e.g., moving an object only a short distance sometimes resulted in an unexpectedly large change in loudness. These observations highlight how individual expectations of sound intensity and distance often vary due to distance compression biases \cite{farnell2010designing, finnegan2016distance}, where the weight of the HMD itself can be a contributing factor to this bias \cite{hmd2009}. Although the mutisensory integration of both visual and auditory stimuli has been proven to gauge better distance localization due to gain re-balancing between senses \cite{odegaard2015biases, crossmodal2016chi}, it has been also observed that in scenarios of level/size discrepancy, the audio component takes precedence over visuals \cite{crossmodal2016chi}, up to the extent that spatial audio, in particular, can overpower visual perception \cite{correa2023spatial, nonspeech1994}. Therefore, when dealing with spatial audio one has to be careful to not overdo with audio. Several participants noted the benefit of having \enquote{visual boundaries} (i.e. weighting on the visual component) in order to have some guidance on level control. 
(P27 \enquote{\emph{Knowing the boundaries of when an object became visible/invisible aka audible/inaudible. Depending on where I was standing in the room, I was wishing there were some guidance on how far I could push out an object on each $XYZ$ plane before they became inaudible/not part of the mix}}). 
Previous research has shown that visual boundaries can provide better sensitivity in spatial localization \cite{boundaries2012}. Additionally, our participants re-iteratively requested a quick way to enable or silence specific virtual audio objects so that they can isolate sounds or compare before-and-after placement. This can be linked to selective attention mechanisms \cite{farnell2010designing}. In particular, the \enquote{cocktail party effect} describes this selective attention phenomenon, where listeners can concentrate on one channel in a noisy environment and dynamically switch attention as needed \cite{nonspeech1994, shelvock2016gestalt}. In a mixing context, muting or soloing audio objects leverages \emph{figure-ground organization} -- also rooted in Gestalt principles -- allowing foreground sounds to be distinguished from background layers. 
This is where the quantitative results of spread difference between scenes start making sense: in the cinematic the spatial spread of audio objects is significantly larger than in the music mixing scene, and this can be attributed to the fact that participants had a bias towards considering cinematic audio objects as more ambient-related and tried to magnify the spread each point-source audio object by extending the sonic perimeter of their mix. In contrast, in the music mixing scene, most of the audio objects were considered as \enquote{salient} and were required to be loud enough in the mix to contribute its timbral and rhythmic component, which lead to narrower spread, mostly at the hand's reach.

Beyond loudness, participants requested control over other audio attributes, such as reverberation levels or pitch. Even though loudness is considered a primary cue for distance estimation \cite{finnegan2016distance}, studies report that reverberation and timbral qualities (e.g., the frequency spectrum) are some other cues that greatly influence distance perception. Participants expected to manipulate these cues in tandem with loudness, similarly, modifying the physical placement of virtual markers for a coherent and holistic mixing experience. There is a lot of previous work relating size of visual markers to pitch \cite{karla2010vision, farnell2010designing, nonspeech1994}, vertical positioning to pitch \cite{bernstein1971, crossmodal2016chi, karla2010vision}, even marker color to pitch \cite{o2024sound, karla2010vision}, some of which converged into heuristics adopted in practice by audio mixing practitioners \cite{gibson1997art, dewey2024}. 

Another recurring feature request was replicating and grouping virtual audio objects. One participant described how they wanted to duplicate a single virtual audio object to occupy various positions, increasing its spread. 
Others recommended grouping virtual audio objects and moving them as a unit or \enquote{layer}, distilling core design principles approaches to layering and soundscape arrangement \cite{farnell2010designing}. Grouping resonates again with Gestalt principles, where similar audio sources -- especially, those sharing timbral or rhythmic traits, are naturally clustered together, evidenced by the similarity and common-fate principles \cite{farnell2010designing, shelvock2016gestalt}. It is of extreme interest how our participants positioned virtual audio objects in the space, forming semantic patterns between pairs or groups of sounds, as it can be observed in the heat map analysis (Subsection \ref{heatmaps}). This is especially evident in the cinematic scene, where the task was to create a compelling sonic environment which would resemble a real scene.
By supporting object duplication, grouping, and spatial layering, AudioMiXR can help organizing granular virtual audio objects into a coherent and sensible whole more effectively.

    \subsection{Quantitative Comparison: Usability, Workload, and Creativity}

To validate and expand on the design lessons from our exploratory study, we conducted a follow-up comparison study (Study 2) contrasting the 6DoF AudioMiXR interface with a traditional 2D panner UI. This allowed us to directly assess whether the qualitative affordances participants described in Study 1, such as embodied interaction, spatial awareness, and dynamic audio feedback translated into measurable advantages across usability, workload, and creative output.

\textbf{Usability (SUS).} Participants rated the AudioMiXR interface as significantly more usable, with improvements across all SUS subscales (Holm-corrected, $p < 0.05$). They found the AR interface to be less cumbersome and reported greater confidence in completing the mixing tasks. This aligns with earlier feedback on proprioception-driven interaction and perceived control, reinforcing the idea that embodied interfaces can enhance usability by reducing cognitive translation between intent and action. These results suggest that even for non-expert users, a well-designed 6DoF interface can exceed the usability of a familiar 2D control scheme.

\textbf{Workload (NASA-TLX).} We observed significantly lower mental demand, frustration, and perceived performance difficulty when participants used AudioMiXR compared to the 2D panner (Holm-corrected $p < 0.05$). These results suggest that the spatial interaction model of AudioMiXR helped reduce the subjective cognitive and emotional load associated with the task, likely by externalizing spatial reasoning and offering more intuitive control. Although other dimensions like effort, temporal demand, and physical demand trended favorably toward the AR interface, they did not reach statistical significance after correction.

\textbf{Creative Output.} We observed a statistically significant improvement across all subscales of our creativity metrics for the AR condition (Holm-corrected $p < 0.0001$), including originality, elaboration, and expressiveness. These findings support our claim that spatial freedom and bodily agency facilitates divergent thinking and experimental design behavior. In particular, users produced more varied spatial arrangements, including novel object groupings and creative layering approaches, echoing qualitative feedback from Study 1 on the importance of gestural expressiveness and environmental awareness.

Taken together, these results substantiate core design lessons identified in our initial study: that AR interfaces leveraging full-body interaction and real-world spatiality can meaningfully enhance the sound design experience—not only in terms of subjective preference or novelty, but also in concrete usability, workload, and creativity metrics. Study 2 thus provides empirical grounding to our earlier design lessons and helps delineate contexts where 6DoF AR tools may be most advantageous, such as spatially rich or exploratory mixing tasks.

\subsection{Future Work and Limitations}

AudioMiXR can serve a wide range of sound design tasks by virtue of its object-based workflow 6DoF AR interface. While our current implementation focuses on AR, the system design is equally adaptable to VR and we discuss this next, along with an outline of key avenues for expanding AudioMiXR's functionality and addressing its current limitations. 

\subsubsection{Extension to VR}

\begin{figure}[ht!]
    \centering
    \includegraphics[width=1.0\linewidth]{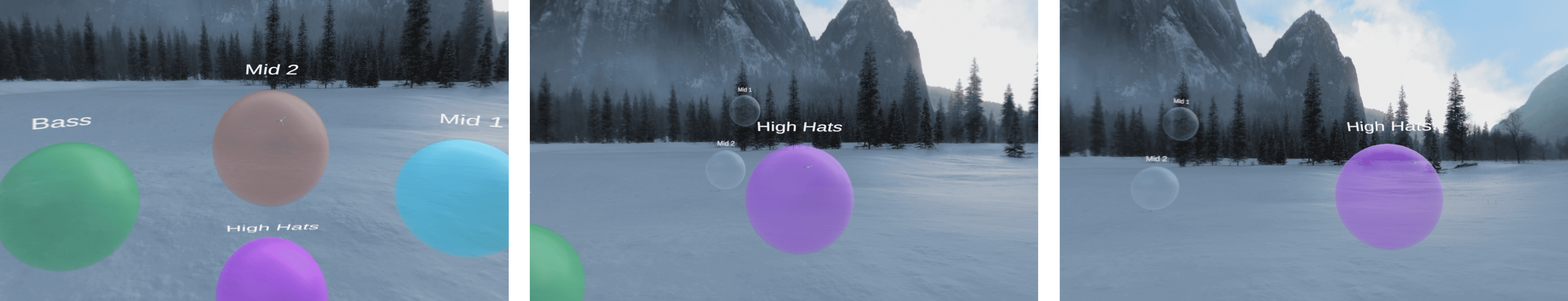}
    \caption{Audio mixing in virtual environments. This example illustrates a demo of AudioMiXR running on top of AVP's native \emph{Environments} application, displaying the virtual Yosemite environment.}
    \label{fig:vr-mixing}
\end{figure}

Beyond AR, extending AudioMiXR into VR offers immersive sound design in fully virtual environments. 
Deploying the same 6DoF interaction model in VR would let users build and experience soundscapes in virtual worlds -- useful for VR cinematic productions, interactive media, and game audio production. However, VR introduces unique challenges, such as cybersickness and occlusion of the real environment, calling for careful design of user feedback and locomotion loops. See Figure \ref{fig:vr-mixing} where we test AudioMiXR within AVP's \emph{Environments} application.

\subsubsection{Multi-user collaboration}

\begin{figure}[ht!]
    \centering
    \includegraphics[width=1.0\linewidth]{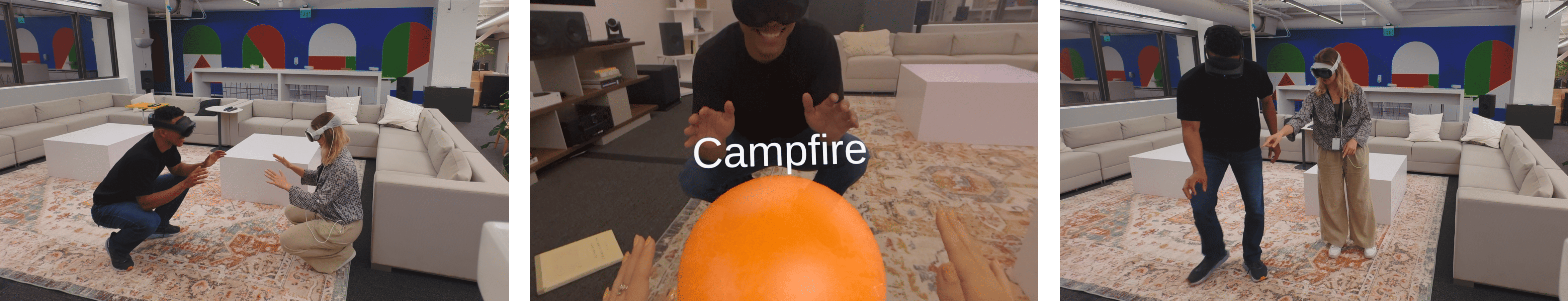}
    \caption{Synchronous multi-user collaborative mixing.}
    \label{fig:collaborative-mixing}
\end{figure}

Although our prototype focuses on single-user, shared XR mixing session is a very attractive direction of research. Enabling synchronous multi-user collaboration in 3D environments, supporting interactions in both remote and co-located spaces, would allow multiple users wearing headsets to co-edit an audio scene in real time. In co-located setups, where users would share the same space, upon AudioMiXR initialization, users would be able to freely navigate in the \emph{XR origin} position within the space and reset the scene to return to the origin. The reset would transform virtual audio objects present in the scene relative to their reset position. Resetting the scene would require verbal communication to ensure all users agree and maintain spatial awareness, as changes would affect everyone sharing the workspace (Figure \ref{fig:collaborative-mixing}). Conversely, in remote collaborations across different physical locations, resetting the workspace would adjust all scene assets, including peer users, relative to the new position. To increase co-presence and spatial awareness in remote AR mixing sessions, visual cues would be used. Each user would be represented by a full-body avatar within the shared workspace, facilitating embodiment and identification. During user-object interactions, such as selection and manipulation, the system would display the user’s hands and highlight the manipulated object. Simultaneous selection and manipulation of the same 3D audio object by multiple users would be prevented to maintain interaction integrity.

\subsection{Applications}

 Our experiments relied on synthesized spatial audio based on generic HRTFs. We invite future studies to couple AudioMiXR with personalized HRTFs, which can enhance even further the localization accuracy of the virtual audio objects. Furthermore, AudioMiXR's design is not restricted to binaural rendering; it could be easily extended to loudspeaker stereophony, allowing users to design mixes that target specific speaker arrays -- opening up research venues centered on immersion and loudspeaker setups.

\subsubsection{Immersive music production.} 
\begin{figure}[ht!]
    \centering
    \includegraphics[width=1.0\linewidth]{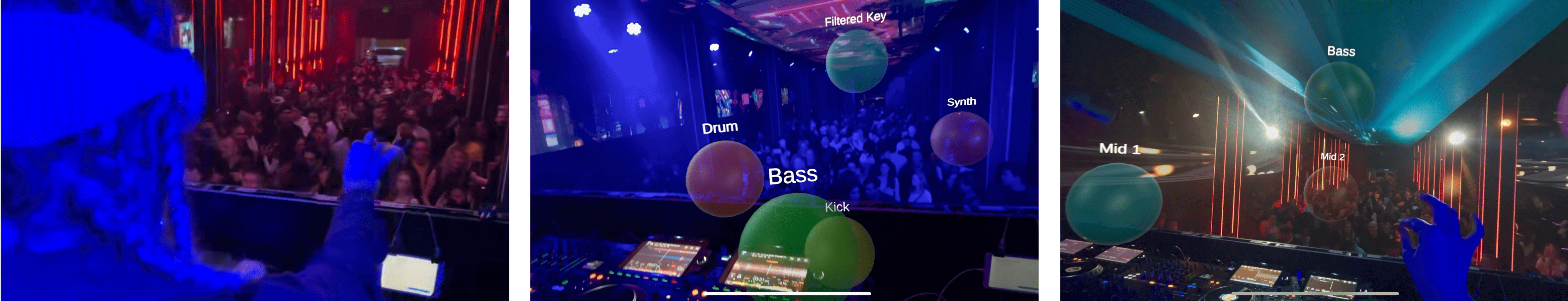}
    \caption{Application scenario illustrating how AudioMiXR could be applied during a live DJ performance. AudioMiXR being used during a DJ set, illustrating how multiple spatial audio objects can be visually placed and manipulated above the turntables and out into the audience. By leveraging 6DoF interaction, DJs can \enquote{grab}, rearrange, and adjust each virtual audio object in real time.}
    \label{fig:applications-3}
\end{figure}

We foresee AudioMiXR as a tool for musicians and producers to be used jointly with DAWs by serving as an 6DoF panner for accurate positioning of spatial audio. In live performances (Figure \ref{fig:applications-3}) --  musicians or DJs could experiment with spatial placement of audio elements while on stage, adjusting soundscapes in real time to match the dynamics of the crowd.

\begin{figure}[ht!]
    \centering
\begin{subfigure}{1.0\linewidth}    \includegraphics[width=1.0\linewidth]{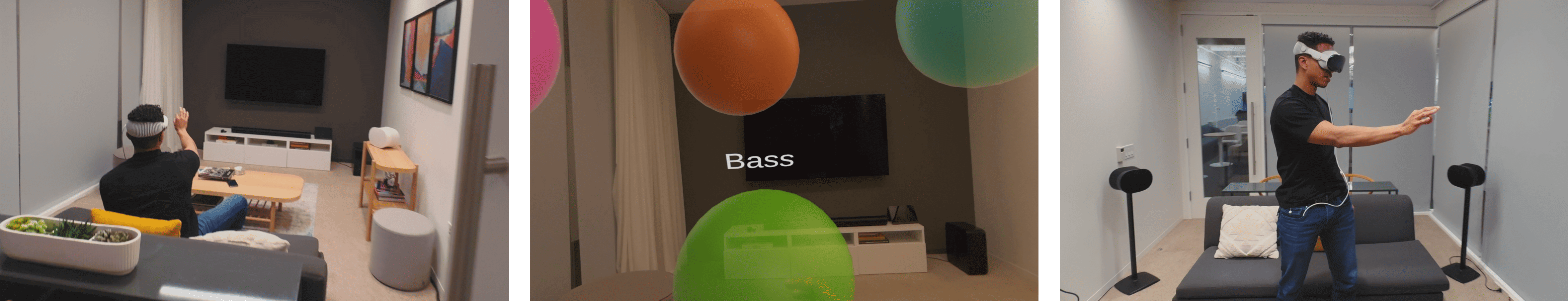}
\end{subfigure}\\
\vspace{1mm}
\begin{subfigure}{1.0\linewidth}
    \centering
    \includegraphics[width=1.0\linewidth]{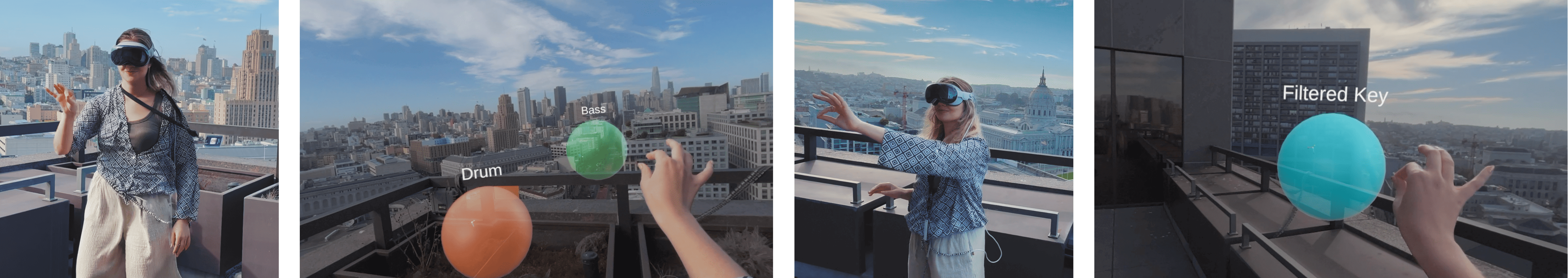}
\end{subfigure}
    \caption{Snapshots of AudioMiXR users in different environments--AudioMiXR allows for audio mixing in any physical environment.}
    \label{fig:applications-1}
\end{figure}

\begin{figure}[ht!]
    \centering
    \includegraphics[width=1.0\linewidth]{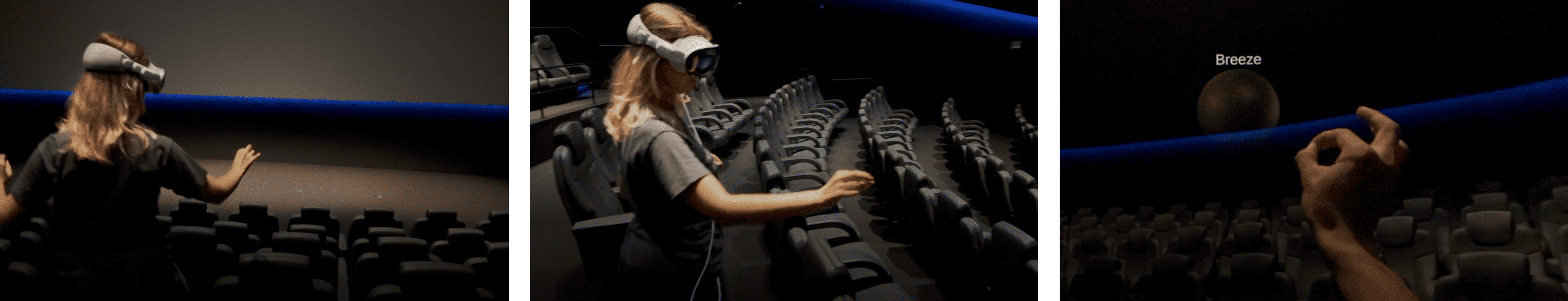}
    \caption{Application scenario illustrating how AudioMiXR could be applied for cinema‑style mixing context. The user positions virtual audio objects throughout the auditorium to explore how spatial cues could be balanced for audiences at different seats.} 
    \label{fig:applications-2}
\end{figure}

\subsubsection{Cinematic and post‑production sound.}
This speculative use case envisions AudioMiXR as a research probe for professional post‑production. Re‑recording teams could position sound effects and dialogue within a cinema, theatre, or on‑set mock‑up, then \textit{walk the room} to verify spatial cues from multiple audience perspectives, reducing reliance on heuristic ``sweet‑spot" listening. Situated authoring of this kind would allow producers to refine object placement in situ, maximizing narrative and emotional impact (Figure~\ref{fig:applications-2}). This probe invites controlled comparison studies with expert mixers (e.g., evaluating localization accuracy, mix completion time), and creative satisfaction against conventional workflows and thereby advances the broader investigation of embodied spatial‑audio tools for cinematic sound.

\subsubsection{Speculative live‑performance scenario}
Figure~\ref{fig:applications-3} sketches a potential use case in which a DJ employs AudioMiXR’s 6DoF controls to place and sculpt virtual audio objects above the turntables and out into the audience, enabling expressive, real‑time spatial mixing.  
We present this vignette as a research probe rather than as a capability of the current prototype: it surfaces open design questions about embodied control in high‑tempo, performance contexts. Positioning AudioMiXR in this way connects our work to broader research venues on immersive authoring tools, creative‑workflow support, and collaborative performance environments, and it outlines a concrete agenda for future investigations into spatial‑audio interaction.

\section{CONCLUSION}


In this work, we presented AudioMiXR, to the best of the authors' knowledge, the first 6DoF AR interface for sound design, enabling direct free-hand manipulation of spatial audio. Therefore, no guidelines for sound design in XR currently exist. We developed two design lessons: (1) Proprioception for AR Sound Design and (2) Balancing Audio-Visual Modalities in AR GUI. The first design lesson underscores how utilizing the body when mixing audio supports perceptibility and self-localization in relation to both the physical environment and the manipulated audio objects. The second design lesson highlights the importance of carefully balancing audio and visual modalities to foster a cohesive mixing experience and provide more intuitive representations of sound.

To substantiate these lessons, we conducted a follow-up comparison study that quantitatively evaluated AudioMiXR against a 2D panner interface. This study confirmed that our AR interface significantly improved usability (SUS), reduced frustration and mental workload (NASA-TLX), and enhanced creative output across multiple dimensions. These results empirically reinforce the qualitative insights from our first study, demonstrating that 6DoF interaction not only feels more engaging and intuitive but also offers measurable benefits in terms of user experience and creative efficacy.

Overall, our findings suggest that embracing 6DoF interaction in XR audio mixing can significantly improve usability, reduce cognitive burden, and expand creative possibilities with spatial audio across domains such as music production, cinematic soundscapes, and interactive media.

\section{Acknowledgements}
We would like to thank the staff at Dolby Laboratories for their continuous support and guidance throughout this research.

\newpage
\bibliographystyle{ACM-Reference-Format}
\bibliography{bibliography}
\newpage
\appendix
\label{sec:appendix}
\section{Survey for professional audio mixers}\label{formative-interview}
Questions for professional audio mixers:
\begin{enumerate}[label=(\arabic*)]
\item Can you describe your general process when mixing audio tracks with spatial audio objects?
\item What tools or software do you typically use for spatial audio localization, mixing, and panning? How do these tools support your workflow?
\item How do you approach spatialization of audio objects within a 3D environment? What are your considerations when positioning audio objects spatially?

\item Can you discuss any specific techniques or strategies you employ to achieve optimal spatialization and panning effects in your projects?
\item In your experience, what are the main challenges or difficulties you encounter when working with spatial audio? How do you typically address these challenges?

\item How often do you collaborate with others (e.g., other sound engineers, content creators) during the spatial audio production process? Can you describe how collaboration influences your workflow and decision-making?

\item Have you encountered any notable successes or breakthroughs in your approach to spatial audio that you'd like to share? How did these impact your projects or clients?

\item Looking ahead, what advancements or improvements do you envision in spatial audio technology or tools that would benefit your work?

\item Can you describe a recent project where spatial audio played a critical role? What were the specific challenges and successes in that project?
\item How do you evaluate the effectiveness of spatial audio mixes? Are there specific criteria or metrics you use to assess quality?
\item What audio attributes do you think can benefit from spatialized interaction? (eg. in a audio panner to manipulate, volume, reverb, equalization etc.)
\item Finally, based on your experience, what advice would you give to someone new to spatial audio mixing and localization?
\end{enumerate}

\section{Demographic Questions}\label{demographic-questions}
\begin{enumerate}[label=(\arabic*)]
\item Age
\item Sex (M/F)
\item Glasses prescription
\item Experience with DAWs (1 = no experience, 2 = beginner, 3 = intermediate, 4 = advanced, 5 = expert)
\item Experience with Augmented reality prior to the experiment 1-5 (1 = no experience, 2 = beginner, 3 = intermediate, 4 = advanced, 5 = expert)
\end{enumerate}

\section{NASA TLX Questions for both non-experts and experts}\label{nasa-tlx}

\begin{enumerate}[label=(\arabic*)]
\item \textbf{Mental Demand}: Using a scale from 1 to 7, where 1 means very low workload and 7 means very high workload, please rate the mental workload involved in manipulating virtual audio objects in the XR system. How mentally demanding was the task?  (1 = very low, 7 = very high)

\item \textbf{Physical Demand}: How physically demanding was it to interact with virtual audio objects using free-hand gestures? How physically demanding was the task? (1 = very low, 7 = very high)

\item \textbf{Temporal Demand}: How hurried or rushed was the pace of the task? (1 = very low, 7 = very high)

\item \textbf{Performance}: How accurately were you able to position and manipulate virtual audio objects using free-hand gestures in the XR environment? How successful were you in accomplishing the task? (1 = perfect, 7 = failure)

\item \textbf{Effort}: How hard did you have to work to accomplish your level of performance? (1 = very low, 7 = very high)

\item \textbf{Frustration}: To what extent did you experience frustration or confusion while using the XR system to manipulate audio objects? How insecure, discouraged, irritated, stressed, and annoyed were you? (1 = very low, 7 = very high)

\item \textbf{Efficiency}: Rate the efficiency of completing tasks (e.g., moving, resizing, rotating) with virtual audio objects in the XR system. (1 = very low, 7 = very high)

\item \textbf{Overall Satisfaction}: On a scale from 1 to 7, how satisfied are you with the overall usability of the XR system for manipulating audio objects? (1 = very low, 7 = very high)
\end{enumerate}

\section{Short-answer survey}\label{short-answer-survey}
Questions for non-experts:
\begin{enumerate}[label=(\arabic*)]
\item How intuitive was the process of navigating and interacting with virtual audio objects using free-hand manipulation?
\item Did you encounter any difficulties or frustrations while attempting to manipulate the virtual audio objects? Please describe.
\item Were there any features or functionalities that were unclear or difficult to understand during your interaction with the system?
\item How comfortable did you feel using the XR system to manipulate audio objects in your physical environment?
\item What aspects of the system did you find most useful or enjoyable? Why?
\item Were there any specific improvements or additional features you would suggest to enhance the usability of the system?
\item (Experts only) Where can you see this interface being applied to your audio mixing workflow?
\end{enumerate}

\end{document}